# Equilibrium and stiffness study of clustered tensegrity structures with the consideration of pulley sizes

Shuo Ma[a,b,1], Yiqian Chen[a,2], Muhao Chen[c,*,3], Robert E. Skelton[c,4]

[a] *College of Civil Engineering, Zhejiang University of Technology, Hangzhou, Zhejiang 310023, China*

[b] *Key Laboratory of Space Structures of Zhejiang Province, Hangzhou, Zhejiang 310058, China*

[c] *Department of Aerospace Engineering, Texas A & M University, College Station, Texas 77840, United States*

## Abstract

This paper presents the equilibrium and stiffness study of clustered tensegrity structures (CTS) considering pulley sizes. We first derive the geometric relationship between clustered strings and pulleys, where the nodal vector is chosen as the generalized coordinate. Then, the equilibrium equations of the clustered tensegrity structure with pulleys based on the Lagrangian method are given. Since the stiffness of a structure is usually weakened when using clustering strings, we formulate the tangent stiffness matrix equations for analysis. It is also shown that as pulley sizes go to zero, the governing equations of the clustered tensegrity system with pulleys yield to the classical clustered tensegrity structure without pulleys, which is consistent with the existing literature. Three examples are demonstrated to validate the given theory. The proposed method allows one to conduct equilibrium, stiffness, and robustness studies of cluster tensegrity structures with pulleys. Nevertheless, the approach developed in this paper is not limited to the tensegrity structures. It can also be applied to a wide range of applications with pulley-rope systems, such as drilling rigs, ocean platform anchors, and cargo cranes.

## 1. Introduction

Tensegrity structures have shown many advantages in designing art pieces [1,2], civil structures [3-5], robotics [6-9], and metamaterials [10-12]. The abundant number of strings in the tensegrity structures has provided many equilibrium states [13], freedom in designing control laws [14,15], tunable geometries for various loading conditions [16,17], and efficient loading path planning approaches [18,19]. However, the adequate strings may also bring much cost in sensors, actuators, and related devices, difficulty in manufacturing, and complexity in the assembly process. An excellent way to moderate these issues is to use clustered strings. A clustered string combines a group of individual adjacent strings into a continuous one that runs over pulleys at the nodes [20].

---

*Corresponding author. Tel.: +1 979-985-8285. muhaochen@tamu.edu (M. Chen)

ORCID(s): 0000-0003-3789-2893 (S. Ma); 0000-0001-9412-5338 (Y. Chen); 0000-0003-1812-6835 (M. Chen); 0000-0001-6503-9115 (R.E. Skelton)

[1]Assistant Professor, College of Civil Engineering, Zhejiang University of Technology, Hangzhou, 310014, China.

[2]Graduate Student, College of Civil Engineering, Zhejiang University of Technology, Hangzhou, 310014, China.

[3]Postdoctoral Researcher, Department of Aerospace Engineering, Texas A&M University, College Station, TX, USA.

[4]TEES Eminent Professor, Department of Aerospace Engineering, Texas A&M University, College Station, TX, USA.

In such a way, the number of individual strings can be significantly reduced. That is, we can make the best of tensegrity while using less hardware, which can broaden the application of tensegrities. A few research have demonstrated the advantages of cluster tensegrities. For example, You and Pellegrino [21,22] proposed cable-stiffened pantographic deployable structures, including a triangular mast and a mesh reflector using sliding cables, and studied the configuration and equilibrium of the structure. Rhode-Barbarigos et al. compared the static, dynamic, and deployment of a deployable tensegrity hollow-rope footbridge for clustered and non-clustered actuated cables [23]. Gomez-Jauregui et al. built a deployable double-layer tensegrity grid with a cluster of struts and cables [24]. Chen et al. [25] studied the formulation and application of multi-node sliding cable elements for the analysis of suspen-dome structures. Ma et al. proposed a deployable tensegrity cable dome with a minimal number of clustered strings [26].

Although the clustering strategy may bring many benefits, the issue we should point out is that the equilibrium states of the CTS can be changed, and the stiffness of a structure is usually weakened by using clustered strings [23]. After the clustering strategy, one must check the structure's statics and stiffness and see if the structure is in an equilibrium state and satisfies stiffness requirements. So, we need analytical tools to conduct the equilibrium and stiffness study of the CTS. One may run to commercial CAE software for help. But from our best knowledge till now, the popular FEM software, such as ANSYS and ABAQUS, are not able to conduct the clustered tensegrity study [27]. Let alone their costly price and require much experience in using the software. Recent studies by a few researchers have paved the way for this problem. For example, Moored and Bart-Smith firstly proposed the concept of clustered tensegrity and studied the prestress mechanism and stability of the clustered tensegrity structures [20]. Ali et al. presented the equilibrium study of the clustered tensegrity structures using a modified dynamic relaxation algorithm [28]. Zhang et al. [29,30] developed an efficient finite element formulation for geometrically nonlinear elasto-plastic analyses of classical and clustered tensegrities based on the co-rotational approach. Kan et al. derived the dynamic of clustered tensegrity structures using the positional formulation FEM [31]. Ma et al. proposed a general approach to the dynamics and control of clustered tensegrity structures [32]. Feng et al. performed optimal active control algorithms of energy-based approaches for effective vibration control of clustered tensegrity structures [33]. However, in most of the CTS studies, the pulley sizes are neglected due to the small size of pulleys compared with structure members and simplifying the problem mathematically. And civil engineers often compensate for the consideration of pulley sizes by giving a large safety factor with their engineering experience. But the pulley sizes cannot be ignored for applications with high precision requirements, accurate prestress assignments, and relatively big pulleys, such as cable net space telescopes and robotic hands. Moreover, the string-to-bar connections are typically mathematically modeled as a string node located at the bar node's center. But in practice, there may always exist a bias in the node positions due to imperfect manufacturing, creeping in materials, errors in assembly, etc.

Regarding the study of pulleys, a few pieces of research has been conducted but mainly on the pulley transmission mechanism. To name a few, Ali et al. formulated a FEM approach to the statics and dynamics of CTS that accommodates sliding-induced frictions [34]. Miyasaka et al. presented a method of modeling robotic systems with closed-circuit cable-pulley transmissions considering the cable-pulley network friction model [35]. Zhang et al. established the kinematics of cable-driven parallel robots considering pulley mechanisms [36]. Xue et al. proposed an estimation method of grasping force for surgical robots based on the model of a cable-pulley system and analyzed the transmission characteristics and friction loss of the pulleys [37]. However, most of them focus on compound pulley systems, and the approach to the study of pulleys to the clustered tensegrity systems remains missing. In other words, there is no adequate analytical research that allows one to study and provide an analytical view of the node connections of clustered tensegrity structures. Thus, this paper tries to fill the gap of the CTS considering pulley sizes with a focus on the influence of equilibrium states and stiffness. To this end, we derived a general approach to the equilibrium equations and stiffness analysis method to the clustered tensegrity system with pulleys. This study allows one to conduct not only the statics analysis of cluster tensegrity structures with pulleys but also the studies of the errors from the node connections.

The paper is structured as follows. Section 2 presents the tensegrity and pulley notations. Section 3 derives the geometric relationship between clustered strings and pulleys. Section 4 derives the equilibrium equations based on the Lagrangian method. Section 5 shows the linearized statics equations and the tangent stiffness matrix of the clustered tensegrity structure with pulleys. Section 6 demonstrates three examples, and Section 7 summarizes the conclusions.

# 2. Definition of a clustered tensegrity considering pulley size

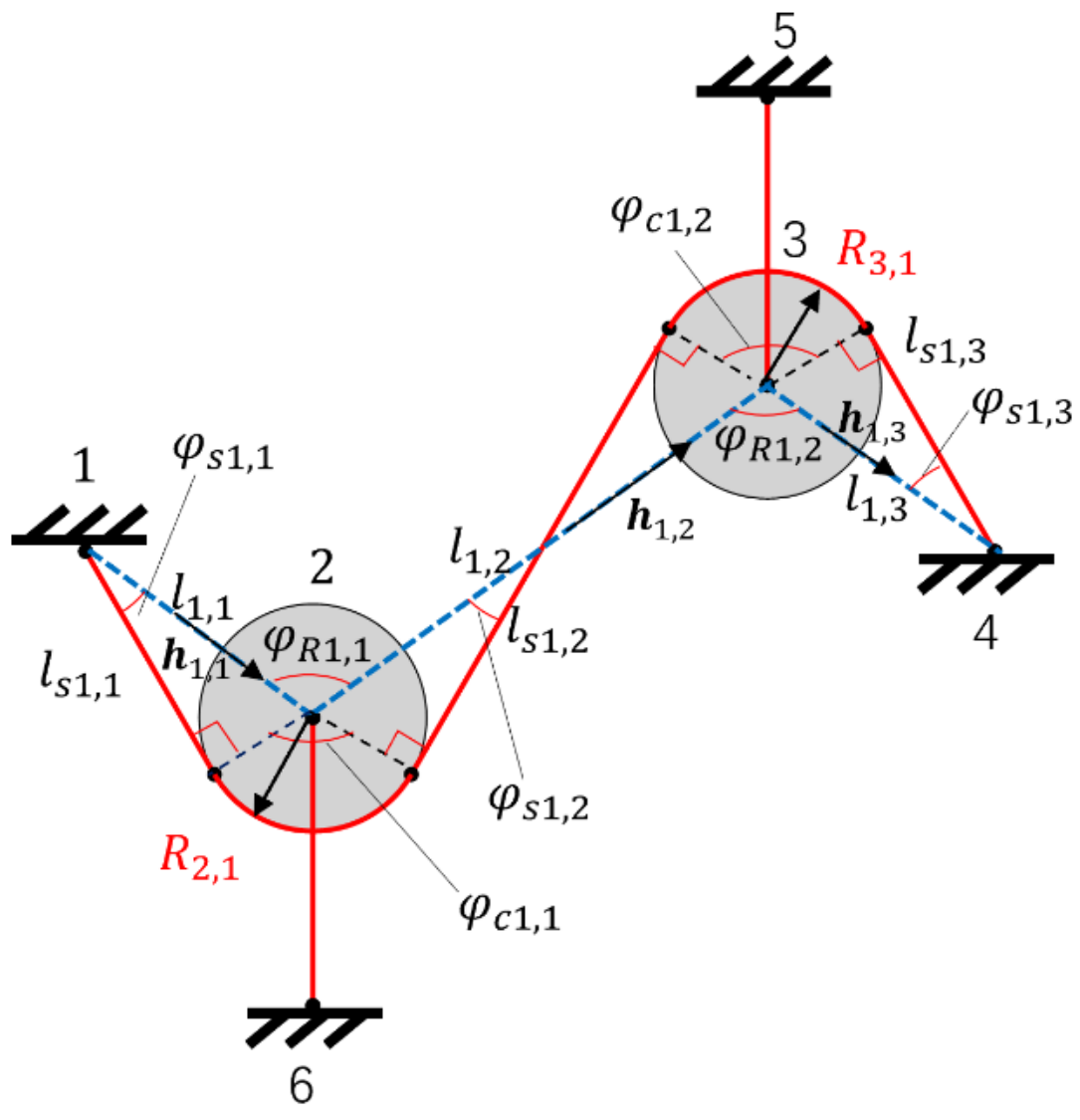

Figure 1 An illustrative example of a clustered tensegrity with pulleys.

## 2.1. Nodal coordinates

There are mainly two connection ways of a string attaching to a joint (directly tied to a node or ring of a pulley), as shown in Figure 1. To distinguish the different connection scenarios, we first define an index system representing the connection of nodes and pulleys. Specifically, we use $\boldsymbol{n}_{i,j}$ to represent the $j$th pulleys on the $i$th ($i = 1, 2, \cdots, n$, $n$ is the total number of nodes.) node. The index $j$ is defined to label the type of a pulleys, where $j = 1$ ($j = 2, \cdots, e_i$, $e_i$ is the number of pulleys) represents a node is pinned to a bar or a string. Though the pulleys in a node can be in different sizes, they are all connected to the same node, $\boldsymbol{n}_{i,1} = \boldsymbol{n}_{i,2} = \boldsymbol{n}_{i,j} = \boldsymbol{n}_{i,e_i} = \boldsymbol{n}_i \in \mathbb{R}^3$ is the same for the nodal coordinate of the $i$th node, and they are repeated in nodal coordinate vector $\boldsymbol{n}$ for the convenience of further computations. There are $n_n$ number of total nodes including the repeated nodes. The nodal coordinate vector $\boldsymbol{n} \in \mathbb{R}^{3n_n}$ of the overall structure network is:

$$\boldsymbol{n} = \left[\boldsymbol{n}_{1,1}, \boldsymbol{n}_{1,2}, \cdots, \boldsymbol{n}_{1,e_1}, \cdots, \boldsymbol{n}_{i,j}, \cdots, \boldsymbol{n}_{n,e_n}\right]^T, \tag{1}$$

where $i = 1,2,\cdots,n$ represents there are $n$ nodes, $j = 1,2,\cdots,e_i$ represents there are $e_i$ type of pulleys connected to the $i$th node.

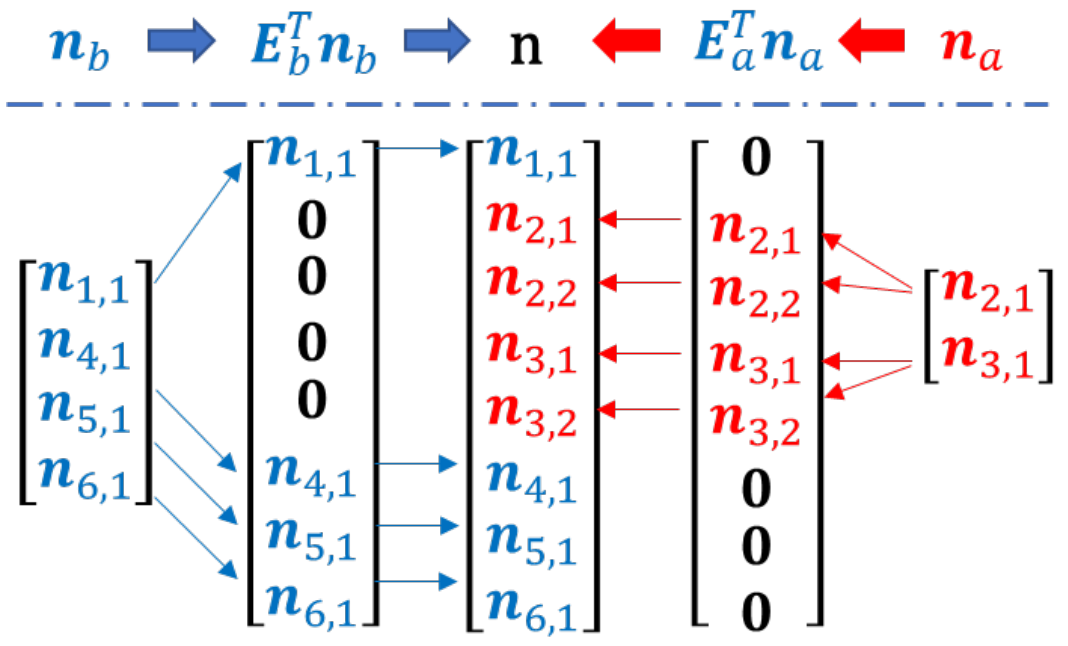

Figure 2 The free and constraint nodal coordinates for Figure 1.

For many practical problems, we need to restrict the motion in specific directions. Fixing the position, velocity, or acceleration of nodes in the structure will reduce the degree of freedom of the structure. We use two index matrices $\boldsymbol{E}_a \in \mathbb{R}^{3n_n \times n_a}$ and $\boldsymbol{E}_b \in \mathbb{R}^{3n_n \times n_b}$ to label the free nodal coordinate vector $\boldsymbol{n}_a \in \mathbb{R}^{n_a}$ and fixed coordinate vector $\boldsymbol{n}_b \in \mathbb{R}^{n_b}$ in the total nodal vector $\boldsymbol{n}$, as shown in Figure 2 for the example given in Figure 1. Noted that $n_a$ and $n_b$ is the number of free and constrained nodal coordinates. And we have the following equation to combine $\boldsymbol{n}_a$ and $\boldsymbol{n}_b$ into $\boldsymbol{n}$:

$$\boldsymbol{n} = \boldsymbol{E}_a\boldsymbol{n}_a + \boldsymbol{E}_b\boldsymbol{n}_b. \tag{2}$$

We have the following equations to extract $\boldsymbol{n}_a$ and $\boldsymbol{n}_b$ from $\boldsymbol{n}$:

$$\boldsymbol{n}_a = \boldsymbol{E}_a^+\boldsymbol{n}, \tag{3}$$

$$\boldsymbol{n}_b = \boldsymbol{E}_b^+\boldsymbol{n}, \tag{4}$$

where $\boldsymbol{V}^+$ is the pseudo-inverse of a matrix $\boldsymbol{V}$.

## 2.2. The radius of the pulleys

The radius of a pulley connected to the $j$th pulley on the $i$th node is noted as $R_{i,j}^n$, we distinguish different radius of the $i$th node because for the string connected to the center of the node, $R_{i,1}^n = 0$, and for strings connected to a pulley, $R_{i,j}^n > 0$. We can further extend the definition of the many pulleys with different radii connected to the same node. The radius vector $\boldsymbol{R} \in \mathbb{R}^{n_n}$ of all the pulleys is:

$$\boldsymbol{R} = \left[R_{1,1}^n \quad R_{1,2}^n \quad \cdots \quad R_{1,e_1}^n \quad \cdots \quad R_{i,j}^n \quad \cdots \quad R_{n,e_n}^n\right]^T. \tag{5}$$

## 2.3. Connectivity matrix

The bar and string topology can be given by the index of its end nodes, labeled by a connectivity matrix $\boldsymbol{C} \in \mathbb{R}^{n_e \times n_n}$, where $n_e$ is the total number of the segments (bars and strings from node to node) in the structure. Suppose the $i$th segment is from node $j$ and to node $k$. The definition of connectivity matrix $\boldsymbol{C}$ ($[\boldsymbol{C}]_{im}$ is the $i$th row and $m$th column) is:

$$[\boldsymbol{C}]_{im} = \begin{cases} -1 & m = j \\ 1 & m = k. \\ 0 & m = \text{else} \end{cases} \tag{6}$$

We also define a start-node connectivity matrix $\boldsymbol{C}_{\text{start}} \in \mathbb{R}^{n_e \times n_n}$ to locate the start points of all the segments:

$$[\boldsymbol{C}_{start}]_{im} = \begin{cases} 1 & m = j \\ 0 & m = \text{else} \end{cases}. \tag{7}$$

Similarly, the end-node connectivity matrix $\boldsymbol{C}_{end} \in \mathbb{R}^{n_e \times n_n}$ is:

$$[\boldsymbol{C}_{end}]_{im} = \begin{cases} 1 & m = k \\ 0 & m = \text{else} \end{cases}. \tag{8}$$

The above three connectivity matrices have the following relation:

$$\boldsymbol{C} = \boldsymbol{C}_{end} - \boldsymbol{C}_{\text{start}}. \tag{9}$$

## 2.4. Side vector of strings in a pulley

For a planar case, we distinguish the left and right sides of a pulley because strings can go through a pulley from either side. Let us use $\mu_{i,j}^n$ to indicate the side positions of the string with the $j$th pulley in the $i$th node, 1 for the string is on the right side of the pulley, and -1 for the string on the left side of the pulley:

$$\mu_{i,j}^n = \begin{cases} 0 & \text{connected to the center of the node} \\ -1 & \text{go around the left side of the pulley} \\ 1 & \text{go around the right side of the pulley} \end{cases}. \tag{10}$$

By stacking all the $\mu_{i,j}^n$ in a vector, we obtain the side vector $\boldsymbol{\mu} \in \mathbb{R}^{n_n}$:

$$\boldsymbol{\mu} = \begin{bmatrix} \mu_{1,1}^n & \mu_{1,2}^n & \cdots & \mu_{1,e_1}^n & \cdots & \mu_{i,j}^n & \cdots & \mu_{n_n,e_n}^n \end{bmatrix}^T. \tag{11}$$

To label the relative position of string and pulley of start and endpoint, we also define $\bar{\mu}_{i,j}$ and $\underline{\mu}_{i,j}$ as the side indicator representing the $j$th segment in the $i$th string:

$$\bar{\mu}_{i,j} = \boldsymbol{S}_{T_{i,j}} \boldsymbol{C}_{start} \boldsymbol{\mu}, \tag{12}$$

$$\underline{\mu}_{i,j} = \boldsymbol{S}_{T_{i,j}} \boldsymbol{C}_{end} \boldsymbol{\mu}, \tag{13}$$

where $\boldsymbol{S}_{T_{i,j}}$ is the sequence matrix introduced in Section 2.6.

## 2.5. Clustering matrix

The clustering matrix $\boldsymbol{S} \in \mathbb{R}^{n_{ec} \times n_e}$ represents the clustering information of the structure segments, where $n_{ec}$ is the total number of members after the clustering strategy, including bars, clustered, and non-clustered strings. The rows in $\boldsymbol{S}$ with multiple 1s represent a clustered string, while the rows in $\boldsymbol{S}$ with only a single 1 represent a non-clustered string:

$$[\boldsymbol{S}]_{ij} = \begin{cases} 1 & \begin{matrix} \text{the } i\text{th element} \\ \text{is composed of the } j\text{th segment.} \end{matrix} \\ 0 & \text{otherwise} \end{cases} \tag{14}$$

The force vector of all members $\boldsymbol{t}_c \in \mathbb{R}^{n_{ec}}$ is:

$$\boldsymbol{t}_c = \widehat{\boldsymbol{E}}_c \widehat{\boldsymbol{A}}_c \hat{\boldsymbol{l}}_{0c}^{-1} (\boldsymbol{l}_c - \boldsymbol{l}_{0c}), \tag{15}$$

where $\boldsymbol{E}_c, \boldsymbol{A}_c, \boldsymbol{l}_{0c}, \boldsymbol{l}_c \in \mathbb{R}^{n_{ec}}$ are Young's modulus, cross-sectional area, rest length, and present length of all members. We can use a clustered matrix to transform the force in members $\boldsymbol{t}_c \in \mathbb{R}^{n_{ec}}$ to force in non-clustered segments $\boldsymbol{t} \in \mathbb{R}^{n_e}$ [28,20]:

$$\boldsymbol{t} = \boldsymbol{S}^T \boldsymbol{t}_c. \tag{16}$$

## 2.6. Sequence matrix

In the existing research, the clustered matrix contains the information on which segments are connected to form a clustered string. Still, the information on the sequence of connected segments is lost, which is crucial if the pulley is considered in the geometry. To represent the information of the sequence of segments in a clustered string, a sequence matrix $\boldsymbol{S}_{T_i} \in \mathbb{R}^{n_{eci} \times n_e}$ is proposed to define the sequence of segments in the $i$th clustered string:

$$[\boldsymbol{S}_{T_i}]_{j,k} = \begin{cases} 1 & k \text{ equals the index of } j\text{th segment in } i\text{th string} \\ 0 & k = \text{else} \end{cases}, \tag{17}$$

where the index of $j$th segment refers to the row number in connectivity matrix $\boldsymbol{C}$ corresponding to the $j$th segment, and $n_{eci}$ is the number of segments in the $i$th member. The $i$th row of the clustered matrix $\boldsymbol{S}_i$ is the sum of each column of the sequence matrix $\boldsymbol{S}_{T_i}$:

$$\boldsymbol{S}_i = \boldsymbol{I}_{(1,n_{eci})} \boldsymbol{S}_{T_i}, \tag{18}$$

where $\boldsymbol{I}_{(1,n_{eci})}$ is a matrix of order $1 \times n_{eci}$. We define the sequence matrix of the whole structure $\boldsymbol{S}_T$ by stacking all $\boldsymbol{S}_{T_i}$:

$$\boldsymbol{S}_T = \begin{bmatrix} \boldsymbol{S}_{T_1} \\ \boldsymbol{S}_{T_2} \\ \vdots \\ \boldsymbol{S}_{T_{nec}} \end{bmatrix}. \tag{19}$$

# 3. The geometry of the CTS with pulleys

## 3.1. Member length and its derivation

This part calculates the length of the $j$th segment of the $i$th clustered string. As shown in Figure 1, the vector of the line connecting the center of pulleys corresponding to the $j$th segment of the $i$th member $\boldsymbol{h}_{i,j} \in \mathbb{R}^3$ can be calculated as:

$$\boldsymbol{h}_{i,j} = \left(\boldsymbol{S}_{T_{i,j}} \boldsymbol{C}\right) \otimes \boldsymbol{I}_3 \boldsymbol{n}. \tag{20}$$

The length can be expressed as:

$$l_{i,j} = |\boldsymbol{h}_{i,j}| = \left|\left(\boldsymbol{S}_{T_{i,j}} \boldsymbol{C}\right) \otimes \boldsymbol{I}_3 \boldsymbol{n}\right|. \tag{21}$$

The relation between $l_{i,j}$ and $\boldsymbol{l}$ is:

$$l_{i,j} = \boldsymbol{S}_{T_{i,j}} \boldsymbol{l}. \tag{22}$$

The length vector of the whole structure $\boldsymbol{l} \in \mathbb{R}^{n_e}$ can be calculated as:

$$\boldsymbol{l} = \sum_{i=1}^{n_{ec}} \sum_{j=1}^{n_{ec_i}} \boldsymbol{S}_{T_{i,j}}^T l_{i,j}. \tag{23}$$

The distance of the straight part corresponding to the $j$th segment of the $i$th member is:

$$l_{S_{i,j}}^2 = l_{i,j}^2 - \bar{R}_{i,j}^2, \tag{24}$$

where $\bar{R}_{ij}$ is the radius from the $j$th segment to the $i$th clustered string:

$$\bar{R}_{i,j} = \boldsymbol{S}_{T_{i,j}} \bar{\boldsymbol{R}} = \boldsymbol{S}_{T_{i,j}} (\hat{\bar{\boldsymbol{\xi}}} \boldsymbol{C}_{start} + \hat{\underline{\boldsymbol{\xi}}} \boldsymbol{C}_{end}) \boldsymbol{R}, \tag{25}$$

where $\bar{\boldsymbol{\xi}} \in \mathbb{R}^{n_e}$ and $\underline{\boldsymbol{\xi}} \in \mathbb{R}^{n_e}$ are the sign of the start and end pulley radius to sum the side of the right triangle, and the $\hat{\boldsymbol{v}}$ operator transforms the vector $\boldsymbol{v}$ into a diagonal matrix. In this situation, the direction of both the start and endpoint of the $j$th segment in the $i$th string is the same, $\bar{\xi}_{i,j}$ or $\underline{\xi}_{i,j}$ is -1 if the radius of the start point or the end point is small, respectively. In other cases, $\bar{\xi}_{i,j}$ and $\underline{\xi}_{i,j}$ is always 1:

$$\bar{\xi}_{i,j} = \begin{cases} -1, & \boldsymbol{S}_{T_{i,j}}\boldsymbol{C}_{start}\boldsymbol{R} < \boldsymbol{S}_{T_{i,j}}\boldsymbol{C}_{end}\boldsymbol{R}, \bar{\mu}_{i,j}\, \underline{\mu}_{i,j} > 0 \\ 1, & \text{else} \end{cases}, \tag{26}$$

$$\underline{\xi}_{i,j} = \begin{cases} -1, & \boldsymbol{S}_{T_{i,j}}\boldsymbol{C}_{end}\boldsymbol{R} < \boldsymbol{S}_{T_{i,j}}\boldsymbol{C}_{start}\boldsymbol{R}, \bar{\mu}_{i,j}\, \underline{\mu}_{i,j} > 0 \\ 1, & \text{else} \end{cases}. \tag{27}$$

$\bar{\boldsymbol{R}} \in \mathbb{R}^{n_e}$ is the vector of radius corresponding to all the segments that form a right triangle with the radius of the pulleys on both ends:

$$\bar{\boldsymbol{R}} = (\hat{\bar{\boldsymbol{\xi}}}\boldsymbol{C}_{start} + \hat{\underline{\boldsymbol{\xi}}}\boldsymbol{C}_{end})\boldsymbol{R}. \tag{28}$$

The relation between $l_{Si,j}$ and $\boldsymbol{l}_S$ is:

$$l_{S_{i,j}} = \boldsymbol{S}_{T_{i,j}}\boldsymbol{l}_S. \tag{29}$$

The length of all the segments $\boldsymbol{l}_S \in \boldsymbol{R}^{n_e}$ can be calculated as:

$$\boldsymbol{l}_S = \sum_{i=1}^{n_{ec}} \sum_{j=1}^{n_{ec_i}} \boldsymbol{S}_{T_{i,j}}^T l_{S_{i,j}}. \tag{30}$$

Or simply by:

$$\boldsymbol{l}_S^2 = \boldsymbol{l}^2 - \bar{\boldsymbol{R}}^2. \tag{31}$$

The derivation of the length of the line connecting the center of pulleys and straight-line length of a segment to nodal coordinate is respectively:

$$\frac{\partial l_{i,j}}{\partial \boldsymbol{n}} = l_{i,j}^{-1}\left(\left(\boldsymbol{S}_{Ti,j}\boldsymbol{C}\right)^T \boldsymbol{S}_{Ti,j}\boldsymbol{C}\right) \otimes \boldsymbol{I}_3\boldsymbol{n}, \tag{32}$$

$$\frac{\partial l_{Si,j}}{\partial \boldsymbol{n}} = l_{Si,j}^{-1}\left(\left(\boldsymbol{S}_{Ti,j}\boldsymbol{C}\right)^T \boldsymbol{S}_{Ti,j}\boldsymbol{C}\right) \otimes \boldsymbol{I}_3\boldsymbol{n}. \tag{33}$$

## 3.2. Angles and their derivatives

### 3.2.1. Angles for the straight segments

As shown in Figure 1, the angle between the *j*th segment and the *i*th member can be related by the geometric properties:

$$\sin \varphi_{Si,j} = \frac{\bar{R}_{i,j}}{l_{i,j}}, \tag{34}$$

$$\cos \varphi_{Si,j} = \frac{l_{Si,j}}{l_{i,j}}. \tag{35}$$

From the above two equations, we have:

$$\varphi_{Si,j} = \text{asin}\frac{\bar{R}_{i,j}}{l_{i,j}}. \tag{36}$$

Stack the elements into a vector form, and we have the angle vector for all the straight segments $\boldsymbol{\varphi}_S \in \mathbb{R}^{n_e}$:

$$\boldsymbol{\varphi}_S = \text{asin}\frac{\bar{\boldsymbol{R}}}{\boldsymbol{l}}. \tag{37}$$

Taking the derivative of Eq.(34), we have:

$$\cos \varphi_{Si,j}\frac{\partial \varphi_{Si,j}}{\partial \boldsymbol{n}} = -l_{i,j}^{-3}\bar{R}_{i,j}\left[\left(\boldsymbol{S}_{Ti,j}\boldsymbol{C}\right)^T \boldsymbol{S}_{Ti,j}\boldsymbol{C}\right] \otimes \boldsymbol{I}_3\boldsymbol{n}. \tag{38}$$

Substitute Eq.(35) into Eq.(38), and we have:

$$\frac{\partial \varphi_{Si,j}}{\partial \boldsymbol{n}} = -l_{i,j}^{-2} l_{Si,j}{}^{-1} \bar{R}_{i,j} \left( \left( \boldsymbol{S}_{Ti,j} \boldsymbol{C} \right)^T \boldsymbol{S}_{Ti,j} \boldsymbol{C} \right) \otimes \boldsymbol{I}_3 \boldsymbol{n}. \tag{39}$$

### 3.2.2. Angles for the circular string

As shown in Table 1, the angle between the two dotted lines connecting the center of the pulleys is:

$$\varphi_{Ri,j} = \arccos \frac{-\boldsymbol{h}_{i,j}^T \boldsymbol{h}_{i,j+1}}{|\boldsymbol{h}_{i,j}||\boldsymbol{h}_{i,j+1}|}. \tag{40}$$

However, this expression is not enough to describe the whole range of rotation because the above equation cannot distinguish the ranges of $[0, \pi)$ and $[\pi, 2\pi)$. Refer to Liu and Paulino[38], the following definition is introduced to expand the domain of our formulation to $[0,2\pi)$:

$$\varphi_{Ri,j} = \eta_{i,j} \arccos \frac{-\boldsymbol{h}_{i,j}^T \boldsymbol{h}_{i,j+1}}{|\boldsymbol{h}_{i,j}||\boldsymbol{h}_{i,j+1}|} \bmod 2\pi \begin{cases} i = 1,2,\cdots,n_{ec} \\ j = 1,2,\cdots,n_{ec_i} - 1 \end{cases}, \tag{41}$$

where $\eta_{i,j}$ is a sign indicator defined as:

$$\eta_{ij} = \begin{cases} sgn(\boldsymbol{r}_{i,j}^T \boldsymbol{h}_{i,j+1}), & \boldsymbol{r}_{i,j}^T \boldsymbol{h}_{i,j+1} \neq 0 \\ 1, & \boldsymbol{r}_{i,j}^T \boldsymbol{h}_{i,j+1} = 0 \end{cases}, \tag{42}$$

where $\boldsymbol{r}_{ij} \in \mathbb{R}^3$ is the vector perpendicular to $\boldsymbol{h}_{ij}$,

$$\boldsymbol{r}_{ij} = \underline{\mu}_{i,j} \boldsymbol{z} \times \boldsymbol{h}_{ij}. \tag{43}$$

And the illustration of the sign indicator $\eta_{i,j}$ is shown in Table 1.

Table 1 Illustration of the sign indicator $\eta_{i,j}$.

| | $\underline{\mu}_{i,j} = 1$ | $\underline{\mu}_{i,j} = -1$ |
|---|---|---|
| $\eta_{i,j} = 1$ | 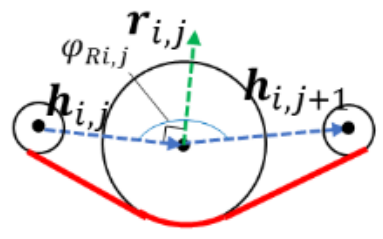 | 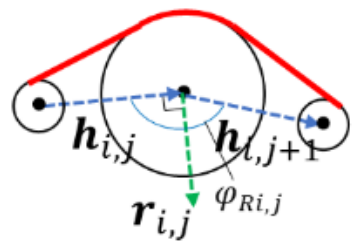 |
| $\eta_{i,j} = -1$ | 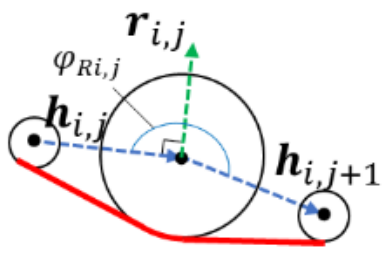 | 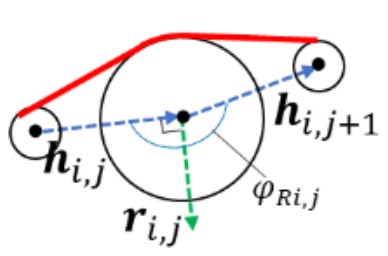 |

Eq.(41) can also be written as:

$$\cos \varphi_{Ri,j} = \eta_{ij} \frac{-\boldsymbol{h}_{i,j}^T \boldsymbol{h}_{i,j+1}}{|\boldsymbol{h}_{i,j}||\boldsymbol{h}_{i,j+1}|}. \tag{44}$$

Taking derivate of Eq.(44), we have:

$$\begin{aligned} \frac{\partial \cos \varphi_{R_{i,j}}}{\partial \boldsymbol{n}} &= -\sin \varphi_{R_{i,j}} \frac{\partial \varphi_{R_{i,j}}}{\partial \boldsymbol{n}} \\ &= -\eta_{i,j} \frac{\frac{\partial (\boldsymbol{h}_{i,j}^T \boldsymbol{h}_{i,j+1})}{\partial \boldsymbol{n}} |\boldsymbol{h}_{i,j}||\boldsymbol{h}_{i,j+1}| - \frac{\partial (|\boldsymbol{h}_{i,j}||\boldsymbol{h}_{i,j+1}|)}{\partial \boldsymbol{n}} (\boldsymbol{h}_{i,j}^T \boldsymbol{h}_{i,j+1})}{|\boldsymbol{h}_{i,j}|^2 |\boldsymbol{h}_{i,j+1}|^2}. \end{aligned} \tag{45}$$

The simplified result of Eq.(45) is:

$$\frac{\partial \varphi_{R_{i,j}}}{\partial \boldsymbol{n}} = \underline{\mu}_{i,j} \left[ \left( \boldsymbol{s}_{T_{i,j}} \boldsymbol{l} \right)^{-2} \left( \boldsymbol{s}_{T_{i,j}} \boldsymbol{c} \right)^T \otimes \boldsymbol{I}_3 \boldsymbol{z} \times \left( \boldsymbol{s}_{T_{i,j}} \boldsymbol{c} \right) \otimes \boldsymbol{I}_3 \boldsymbol{n} - \left( \boldsymbol{s}_{T_{i,j+1}} \boldsymbol{l} \right)^{-2} \left( \boldsymbol{s}_{T_{i,j+1}} \boldsymbol{c} \right)^T \otimes \boldsymbol{I}_3 \boldsymbol{z} \times \left( \boldsymbol{s}_{T_{i,j+1}} \boldsymbol{c} \right) \otimes \boldsymbol{I}_3 \boldsymbol{n} \right]. \tag{46}$$

The derivation of Eq.(46) is given in Appendix A.

## 3.3. The start and end nodal coordinates of the straight strings

The nodal coordinate of the start node $\overline{\boldsymbol{n}}_{i,j} \in \mathbb{R}^3$ and end node $\underline{\boldsymbol{n}}_{i,j} \in \mathbb{R}^3$ of the straight strings of the $j$th segment in the $i$th member can be calculated by the geometric information as shown in Figure 3.

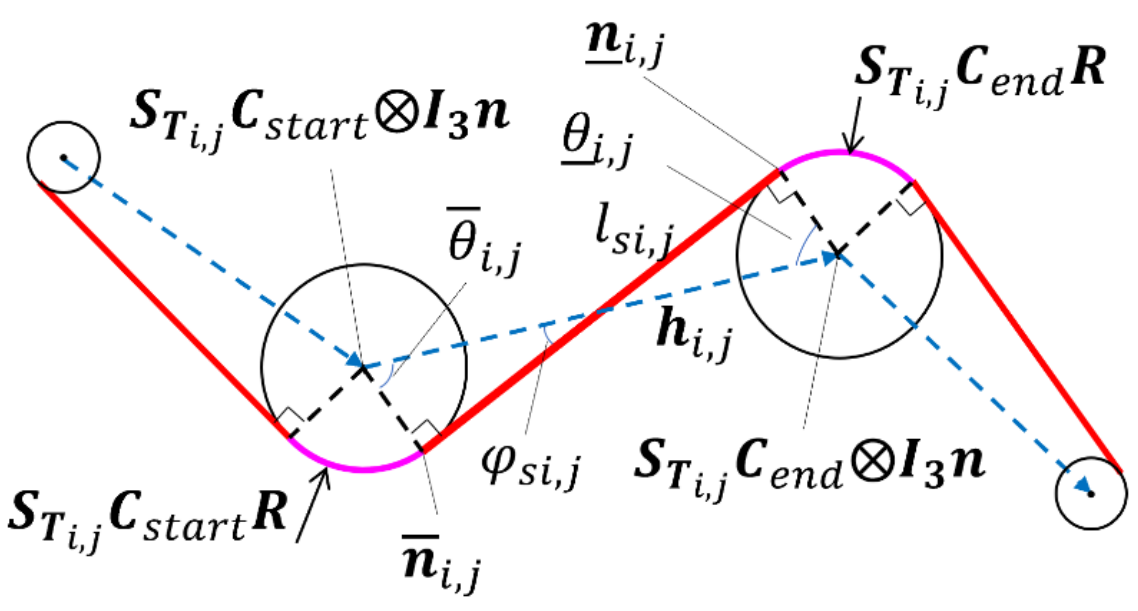

Figure 3 A diagram for the start and end nodes of the straight strings. (Red solid lines are the straight strings, solid pink lines are the circular strings, and blue dotted lines connect the pulley centers.)

The nodal coordinates of the start node and end node of the $j$th segment of the $i$th member are:

$$\overline{\boldsymbol{n}}_{i,j} = \boldsymbol{S}_{T_{i,j}}\boldsymbol{C}_{start} \otimes \boldsymbol{I}_3\boldsymbol{n} + \boldsymbol{S}_{T_{ij}}\boldsymbol{C}_{start}\boldsymbol{R}\overline{\boldsymbol{T}}_{i,j}\frac{\vec{\boldsymbol{h}}_{ij}}{|\boldsymbol{h}_{ij}|}, \tag{47}$$

$$\underline{\boldsymbol{n}}_{i,j} = \boldsymbol{S}_{T_{i,j}}\boldsymbol{C}_{end} \otimes \boldsymbol{I}_3\boldsymbol{n} - \boldsymbol{S}_{T_{ij}}\boldsymbol{C}_{end}\boldsymbol{R}\underline{\boldsymbol{T}}_{i,j}\frac{\vec{\boldsymbol{h}}_{ij}}{|\boldsymbol{h}_{ij}|}, \tag{48}$$

where $\overline{\boldsymbol{T}}_{i,j}$ and $\underline{\boldsymbol{T}}_{i,j} \in \mathbb{R}^{3\times3}$ are the rotation matrices:

$$\overline{\boldsymbol{T}}_{ij} = \begin{bmatrix} \cos\overline{\theta}_{ij} & -\sin\overline{\theta}_{ij} \\ \sin\overline{\theta}_{ij} & \cos\overline{\theta}_{ij} \end{bmatrix}, \tag{49}$$

$$\underline{\boldsymbol{T}}_{ij} = \begin{bmatrix} \cos\underline{\theta}_{ij} & -\sin\underline{\theta}_{ij} \\ \sin\underline{\theta}_{ij} & \cos\underline{\theta}_{ij} \end{bmatrix}. \tag{50}$$

And $\overline{\theta}_{i,j}$ is the angle between the vector connecting the center of the pulley to the start node of the straight segment and $\boldsymbol{h}_{i,j}$ (as shown in Figure 3):

$$\overline{\theta}_{ij} = -\overline{\mu}_{i,j}[\frac{\pi}{2} - S_{T_{i,j}}\boldsymbol{\varphi}_s], \tag{51}$$

$$\underline{\theta}_{ij} = \underline{\mu}_{i,j}[\frac{\pi}{2} - S_{T_{i,j}}\boldsymbol{\varphi}_s]. \tag{52}$$

## 3.4. The total length of the clustered strings

The total length of the $i$th member is:

$$l_{C_i} = \sum_{j=1}^{n_{eci}}\boldsymbol{S}_{T_{i,j}}\boldsymbol{l}_s + \sum_{j=1}^{n_{eci}-1}\boldsymbol{S}_{T_{i,j}}\boldsymbol{C}_{end}\boldsymbol{R}\varphi_{C_{i,j}}, \tag{53}$$

where $\varphi_{C_{i,j}}$ is the angle of the circular string around the pulley:

$$\begin{aligned} \varphi_{c_{ij}} &= 2\pi - \varphi_{R_{ij}} - \left(\frac{\pi}{2} - \underline{\xi}_{i,j}\varphi_{S_{i,j}}\right) - \left(\frac{\pi}{2} - \overline{\xi}_{i,j+1}\varphi_{S_{i,j+1}}\right) \\ &= \pi - \varphi_{R_{ij}} + \underline{\xi}_{i,j}\varphi_{S_{i,j}} + \overline{\xi}_{i,j+1}\varphi_{S_{i,j+1}}. \end{aligned} \tag{54}$$

Substitute Eq.(54) into Eq.(53), and we have:

$$l_{c_i} = \sum_{j=1}^{n_{eci}} \boldsymbol{S}_{T_{i,j}} \boldsymbol{l}_s + \sum_{j=1}^{n_{eci}-1} \left(\boldsymbol{S}_{T_{i,j}} \boldsymbol{C}_{end} \boldsymbol{R}\right)\left(\pi - \varphi_{R_{i,j}}\right)$$
$$+ \sum_{j=1}^{n_{eci}} \left(\boldsymbol{S}_{T_{i,j}} \boldsymbol{C}_{start} \boldsymbol{R} \overline{\xi}_{i,j} \varphi_{S_{i,j}} + \boldsymbol{S}_{T_{i,j}} \boldsymbol{C}_{end} \boldsymbol{R} \underline{\xi}_{i,j} \varphi_{S_{i,j}}\right). \quad (55)$$

# 4. Equilibrium equations

## 4.1. The Lagrangian method

The general form of the Lagrangian equation is:

$$\frac{\mathrm{d}}{\mathrm{d}t} \frac{\partial L}{\partial \dot{\boldsymbol{n}}_a} - \frac{\partial L}{\partial \boldsymbol{n}_a} = \boldsymbol{f}_{npa}, \quad (56)$$

where $L = T - V$ is the Lagrangian function, $T$ and $V$ are the kinetic energy and potential energy of the system, $\boldsymbol{n}_a$ is the minimal coordinate of the system, $\boldsymbol{f}_{npa} \in \mathbb{R}^{n_a}$ is the non-potential force vector exerted on the free nodal coordinate, and its relation with the non-potential force $\boldsymbol{f}_{np}$ exerted on all nodes is:

$$\boldsymbol{f}_{npa} = \boldsymbol{E}_a^T \boldsymbol{f}_{np}. \quad (57)$$

For the statics problem, the kinetic energy $T$ is zero, and the Lagrangian method degenerates to:

$$\frac{\partial V}{\partial \boldsymbol{n}_a} = \boldsymbol{f}_{npa}. \quad (58)$$

Eq.(58) is consistent with the principle of stationary total potential energy and the principle of virtual work. However, using the Lagrangian method to derive the equilibrium equation will make it easy to extend to the future study of the dynamics. According to Eq.(2), the nodal coordinate of the whole structure $\boldsymbol{n}$ contains the information of the free nodal coordinate $\boldsymbol{n}_a$, so the left part of Eq.(58) can be written as:

$$\frac{\partial V}{\partial \boldsymbol{n}_a} = \frac{\partial \boldsymbol{n}^{\mathrm{T}}}{\partial \boldsymbol{n}_a} \frac{\partial V}{\partial \boldsymbol{n}} = \boldsymbol{E}_a^T \frac{\partial V}{\partial \boldsymbol{n}}. \quad (59)$$

Substitute Eq. (57) and(59) into Eq. (58), and we have:

$$\boldsymbol{E}_a^T \frac{\partial V}{\partial \boldsymbol{n}} = \boldsymbol{E}_a^T \boldsymbol{f}_{np}. \quad (60)$$

Also noted that if there are no boundary constraints and the pulley size is not considered, $\boldsymbol{E}_a$ will be an identity matrix, and the Eq. (60) will degenerate to $\frac{\partial V}{\partial \boldsymbol{n}} = \boldsymbol{f}_{np}$.

## 4.2. Energy function

The total potential energy $V$ of the system is composed of strain potential energy $V_e$ and gravitational potential energy $V_g$:

$$V = V_e + V_g. \tag{61}$$

### 4.2.1. Strain potential energy

There is no deformation in the pulley, so the strain potential energy is only stored in the axial members:

$$V_e = \sum_{i=1}^{n_{ec}} \int_{l_{0_{ci}}}^{l_{ci}} \boldsymbol{t}_{ci}\, \mathrm{d}u. \tag{62}$$

Substitute Eq.(55) into Eq. (62), and we have the partial derivative of $V_e$ with respect to $\boldsymbol{n}$:

$$\frac{\partial V_e}{\partial \boldsymbol{n}} = \sum_{i=1}^{n_{ec}} \frac{\partial V_e}{\partial l_{c_i}} \frac{\partial l_{c_i}}{\partial \boldsymbol{n}} = \sum_{i=1}^{n_{ec}} t_{c_i} \left[ \sum_{j=1}^{n_{eci}} \frac{\partial \boldsymbol{S}_{T_{i,j}} \boldsymbol{l}_s}{\partial \boldsymbol{n}} - \sum_{j=1}^{n_{eci-1}} \left( \boldsymbol{S}_{T_{ij}} \boldsymbol{C}_{end} \boldsymbol{R} \right) \frac{\partial \varphi_{R_{ij}}}{\partial \boldsymbol{n}} \right] + \sum_{j=1}^{n_{eci}} \left[ \boldsymbol{S}_{T_{ij}} (\boldsymbol{C}_{start} \boldsymbol{R}) (\boldsymbol{S}_{T_{ij}} \bar{\boldsymbol{\xi}}) \frac{\partial \varphi_{S_{ij}}}{\partial \boldsymbol{n}} + \boldsymbol{S}_{T_{ij}} (\boldsymbol{C}_{end} \boldsymbol{R}) (\boldsymbol{S}_{T_{ij}} \underline{\boldsymbol{\xi}}) \frac{\partial \varphi_{S_{ij}}}{\partial \boldsymbol{n}} \right]. \tag{63}$$

Substitute Eq.(33) into the first part of Eq.(63), we have:

$$\begin{aligned}
\sum_{i=1}^{n_{ec}} t_{ci} \sum_{j=1}^{n_{eci}} \frac{\partial \boldsymbol{S}_{T_{ij}} \boldsymbol{l}_s}{\partial \boldsymbol{n}} &= \sum_{i=1}^{n_{ec}} \sum_{j=1}^{n_{eci}} \left[ \left( \boldsymbol{S}_{T_{ij}} \boldsymbol{C} \right)^T \left( \boldsymbol{S}_{T_{ij}} \boldsymbol{l}_s^{-1} \right) \left( \boldsymbol{S}_{T_{ij}} \boldsymbol{t} \right) \left( \boldsymbol{S}_{T_{ij}} \boldsymbol{C} \right) \right] \otimes \boldsymbol{I}_3 \boldsymbol{n} \\
&= \sum_{i=1}^{n_{ec}} \sum_{j=1}^{n_{eci}} \left[ \left( \boldsymbol{S}_{T_{ij}} \boldsymbol{C} \right)^T \left( \boldsymbol{S}_{T_{ij}} \hat{\boldsymbol{l}}_s^{-1} \hat{\boldsymbol{t}} \boldsymbol{S}_{T_{ij}}^T \right) \left( \boldsymbol{S}_{T_{ij}} \boldsymbol{C} \right) \right] \otimes \boldsymbol{I}_3 \boldsymbol{n} \\
&= [(\boldsymbol{S}_T \boldsymbol{C})^T \boldsymbol{S}_T \hat{\boldsymbol{l}}_s^{-1} \boldsymbol{S}_T^T \boldsymbol{S}_T \hat{\boldsymbol{t}} \boldsymbol{S}_T^T \boldsymbol{S}_T \boldsymbol{C}] \otimes \boldsymbol{I}_3 \boldsymbol{n} \\
&= (\boldsymbol{C}^T \hat{\boldsymbol{l}}_s^{-1} \hat{\boldsymbol{t}} \boldsymbol{C}) \otimes \boldsymbol{I}_3 \boldsymbol{n}.
\end{aligned} \tag{64}$$

Substitute Eqs. (46), (12), (13), and (8) into the second part of Eq. (63), we have:

$$\begin{aligned}
&\sum_{i=1}^{n_{ec}} t_{ci} \sum_{j=1}^{n_{eci}} \left( \boldsymbol{S}_{T_{ij}} \boldsymbol{C}_{end} \boldsymbol{R} \right) \frac{\partial \varphi_{R_{i,j}}}{\partial \boldsymbol{n}} \\
&= \sum_{i=1}^{n_{ec}} \sum_{j=1}^{n_{eci}-1} \left[ \left( \boldsymbol{S}_{T_{ij}} \boldsymbol{C} \right)^T \underline{\mu}_{i,j} \left( \boldsymbol{S}_{T_{ij}} \boldsymbol{C}_{end} \boldsymbol{R} \right) \left( \boldsymbol{S}_{T_{ij}} \hat{\boldsymbol{l}}^{-2} \boldsymbol{t} \right) \boldsymbol{z} \times \left( \boldsymbol{S}_{T_{i,j}} \boldsymbol{C} \right) \right] \otimes \boldsymbol{I}_3 \boldsymbol{n} \\
&\quad - \sum_{i=1}^{n_{ec}} \sum_{j=2}^{n_{eci}} \left[ \left( \boldsymbol{S}_{T_{ij}} \boldsymbol{C} \right)^T \overline{\mu}_{i,j} \left( \boldsymbol{S}_{T_{ij}} \boldsymbol{C}_{start} \boldsymbol{R} \right) \left( \boldsymbol{S}_{T_{ij}} \hat{\boldsymbol{l}}^{-2} \boldsymbol{t} \right) \boldsymbol{z} \times \left( \boldsymbol{S}_{T_{i,j}} \boldsymbol{C} \right) \right] \otimes \boldsymbol{I}_3 \boldsymbol{n} \\
&= \left[ (\boldsymbol{S}_T \boldsymbol{C})^T \boldsymbol{S}_T \left( \widehat{\boldsymbol{C} \boldsymbol{\mu} \boldsymbol{R}} \right) \boldsymbol{S}_T^T \boldsymbol{S}_T \hat{\boldsymbol{l}}^{-2} \boldsymbol{S}_T^T \boldsymbol{S}_T \hat{\boldsymbol{t}} \boldsymbol{S}_T^T \right] \otimes \boldsymbol{I}_3 \boldsymbol{I}_{ne} \otimes [\boldsymbol{z}]^\times (\boldsymbol{S}_T \boldsymbol{C}) \otimes \boldsymbol{I}_3 \boldsymbol{n} \\
&= \left[ \boldsymbol{C}^T \left( \widehat{\boldsymbol{C} \boldsymbol{\mu} \boldsymbol{R}} \right) \hat{\boldsymbol{l}}^{-2} \hat{\boldsymbol{t}} \right] \otimes \boldsymbol{I}_3 \boldsymbol{I}_{ne} \otimes [\boldsymbol{z}]^\times \boldsymbol{C} \otimes \boldsymbol{I}_3,
\end{aligned} \tag{65}$$

where $[\boldsymbol{z}]^\times$ is the screw-symmetric matrix of the vector $\boldsymbol{z}$. Substitute Eq.(39) into the third part of Eq. (63), we have:

$$\sum_{i=1}^{n_{ec}} t_{ci} \sum_{j=1}^{n_{eci}} [\boldsymbol{S}_{T_{ij}}(\boldsymbol{C}_{start}\boldsymbol{R})(\boldsymbol{S}_{T_{ij}}\overline{\boldsymbol{\xi}}) \frac{\partial \varphi_{S_{i,j}}}{\partial \boldsymbol{n}} + \boldsymbol{S}_{T_{ij}}(\boldsymbol{C}_{end}\boldsymbol{R})(\boldsymbol{S}_{T_{ij}}\underline{\boldsymbol{\xi}}) \frac{\partial \varphi_{S_{i,j}}}{\partial \boldsymbol{n}}]$$

$$= -\sum_{i=1}^{n_{ec}} \sum_{j=1}^{n_{eci}} \left[\left(\boldsymbol{S}_{T_{ij}}\boldsymbol{C}\right)^T \boldsymbol{S}_{T_{ij}} \left[\hat{\boldsymbol{l}}^{-2} \hat{\boldsymbol{l}}_s^{-1} \widehat{\boldsymbol{R}} \hat{\boldsymbol{t}} \left(\widehat{\overline{\boldsymbol{\xi}}} \widehat{\boldsymbol{C}_{start}\boldsymbol{R}} + \widehat{\underline{\boldsymbol{\xi}}} \widehat{\boldsymbol{C}_{end}\boldsymbol{R}}\right)\right] \boldsymbol{S}_{T_{ij}}^T \left(\boldsymbol{S}_{T_{ij}}\boldsymbol{C}\right)\right] \otimes \boldsymbol{I}_3\boldsymbol{n} \tag{66}$$

$$= -\left[(\boldsymbol{S}_T\boldsymbol{C})^T \boldsymbol{S}_T \left(\hat{\boldsymbol{l}}^{-2} \hat{\boldsymbol{l}}_s^{-1} \widehat{\boldsymbol{R}} \hat{\boldsymbol{t}} \widehat{\boldsymbol{R}}\right) \boldsymbol{S}_T^T (\boldsymbol{S}_T\boldsymbol{C})\right] \boldsymbol{I}_3\boldsymbol{n}$$

$$= -\left(\boldsymbol{C}^T \hat{\boldsymbol{l}}^{-2} \hat{\boldsymbol{l}}_s^{-1} \hat{\boldsymbol{t}} \widehat{\boldsymbol{R}}^2 \boldsymbol{C}\right) \otimes \boldsymbol{I}_3\boldsymbol{n}.$$

### 4.2.2. Gravitational potential energy

The gravitational potential energy is relative to any member that has mass. In the system, all the axial members and point mass will contribute to gravitational potential energy:

$$V_g = V_{ge} + V_{gp}. \tag{67}$$

The gravitational potential energy corresponding to the axial elements $V_{ge}$ is:

$$V_{ge} = \sum_{i=1}^{n_e} \frac{m_{ei}}{2} [a_x \quad a_y \quad a_z] \begin{bmatrix} x_j^i + x_k^i \\ y_j^i + y_k^i \\ z_j^i + z_k^i \end{bmatrix}$$

$$= \sum_{i=1}^{n_e} \frac{m_{ei}}{2} [a_x \quad a_y \quad a_z] \, |\boldsymbol{C}_i| \otimes \boldsymbol{I}_3\boldsymbol{n} \tag{68}$$

$$= \frac{1}{2} (\boldsymbol{m}_e^T |\boldsymbol{C}|) \otimes [a_x \quad a_y \quad a_z] \boldsymbol{n},$$

where $m_{ei}$ is the mass of the $i$th axial element, and $\boldsymbol{m}_e \in \mathbb{R}^{n_e}$ is the mass vector of all axial elements, $a_x, a_y, a_z$ are the gravitational acceleration in the $X$, $Y$, and $Z-$direction. The gravitational potential energy of the pulleys and point mass $V_{gp}$ is:

$$V_{gp} = \sum_{i=1}^{n_n} m_{pi} \otimes [a_x \quad a_y \quad a_z] \begin{bmatrix} x_i \\ y_i \\ z_i \end{bmatrix} \tag{69}$$

$$= \boldsymbol{m}_p^T \otimes [a_x \quad a_y \quad a_z] \boldsymbol{n}.$$

where $m_{pi}$ is the mass of the $i$th node or pulley, and $\boldsymbol{m}_p \in \mathbb{R}^{n_n}$ is the point mass vector. The partial derivative of $V_g$ to $\boldsymbol{n}$ is:

$$\frac{\partial V_g}{\partial \boldsymbol{n}} = \frac{\partial V_{ge}}{\partial \boldsymbol{n}} + \frac{\partial V_{gp}}{\partial \boldsymbol{n}}$$

$$= \left(\frac{1}{2} |\boldsymbol{C}|^T \boldsymbol{m}_e + \boldsymbol{m}_p\right) \otimes [a_x \quad a_y \quad a_z]^T = \boldsymbol{g}, \tag{70}$$

where $\boldsymbol{g}$ is the gravitational force vector.

## 4.3. Nonlinear equilibrium equation

Substitute Eqs. (61), (64), (65), (66), and (70) into Eq.(60), we have

$$\boldsymbol{E}_a^T\boldsymbol{K}\boldsymbol{n} = \boldsymbol{E}_a^T(\boldsymbol{f}_{np} - \boldsymbol{g}), \tag{71}$$

where $\boldsymbol{K} \in \mathbb{R}^{3n_n \times 3n_n}$ is the stiffness matrix,

$$\boldsymbol{K} = (\boldsymbol{C}^T\hat{\boldsymbol{l}}_S\hat{\boldsymbol{l}}^{-2}\hat{\boldsymbol{t}}\boldsymbol{C}) \otimes \boldsymbol{I}_3 - [\boldsymbol{C}^T(\widehat{\boldsymbol{C}\widehat{\boldsymbol{\mu}}\boldsymbol{R}})\hat{\boldsymbol{l}}^{-2}\hat{\boldsymbol{t}}] \otimes \boldsymbol{I}_3\boldsymbol{I}_{ne} \otimes [\boldsymbol{z}]^{\times}\boldsymbol{C} \otimes \boldsymbol{I}_3. \tag{72}$$

We can observe that when $\boldsymbol{R}$ approaches zero, the tangent stiffness matrix equals $\lim_{\boldsymbol{R}\to 0}\boldsymbol{K} = (\boldsymbol{C}^T\hat{\boldsymbol{l}}^{-1}\hat{\boldsymbol{t}}\boldsymbol{C}) \otimes \boldsymbol{I}_3$. This is consistent with the traditional tensegrity systems (TTS refers to non-clustered without considering pulley sizes) [27]. Substitute Eq. (2) into Eq. (71), and we have:

$$\boldsymbol{K}_{aa}\boldsymbol{n}_a = \boldsymbol{E}_a^T(\boldsymbol{f}_{np} - \boldsymbol{g}) - \boldsymbol{K}_{ab}\boldsymbol{n}_b, \tag{73}$$

where:

$$\boldsymbol{K}_{aa} = \boldsymbol{E}_a^T\boldsymbol{K}\boldsymbol{E}_a, \boldsymbol{K}_{ab} = \boldsymbol{E}_a^T\boldsymbol{K}\boldsymbol{E}_b. \tag{74}$$

Eq. (71) is the reduced-order nonlinear equilibrium equation of the clustered tensegrity considering pulley size. We can observe that the equilibrium equation of clustered tensegrity can quickly degenerate to both clustered tensegrity without considering the pulley size and traditional tensegrity structure without clustered strings.

# 5. Linearized equilibrium equation and tangent stiffness matrix

## 5.1. Linearized equilibrium equation

Using Taylor's expansion, the Eq. (71) about an equilibrium configuration $\boldsymbol{n}^i$ in the $i$th iteration step, we have the linearized equilibrium equation:

$$\boldsymbol{E}_a^T[\boldsymbol{K}|_{\boldsymbol{n}^i}\boldsymbol{n}^i + \boldsymbol{K}_T(\boldsymbol{n}^{i+1} - \boldsymbol{n}^i)] = \boldsymbol{E}_a^T(\boldsymbol{f}_{np} - \boldsymbol{g}), \tag{75}$$

where $\boldsymbol{K}_T \in \mathbb{R}^{3n_n \times 3n_n}$ is the tangent stiffness matrix of the structure, $\boldsymbol{K}|_{\boldsymbol{n}^i}$ is the stiffness matrix in $\boldsymbol{n}^i$ configuration. Substitute Eq. (2) into Eq. (75), and we have:

$$\boldsymbol{K}_{aa}|_{\boldsymbol{n}^i}\boldsymbol{n}_a^i + \boldsymbol{K}_{Taa}(\boldsymbol{n}_a^{i+1} - \boldsymbol{n}_a^i) = \boldsymbol{E}_a^T(\boldsymbol{f}_{np} - \boldsymbol{g}) - \boldsymbol{K}_{ab}|_{\boldsymbol{n}^i}\boldsymbol{n}_b^i - \boldsymbol{K}_{Tab}(\boldsymbol{n}_b^{i+1} - \boldsymbol{n}_b^i), \tag{76}$$

where:

$$\boldsymbol{K}_{Taa} = \boldsymbol{E}_a^T\boldsymbol{K}_T\boldsymbol{E}_a, \boldsymbol{K}_{Tab} = \boldsymbol{E}_a^T\boldsymbol{K}_T\boldsymbol{E}_b. \tag{77}$$

By solving Eq. (76), we can obtain a new configuration $\boldsymbol{n}_a^{i+1}$ in the $i$+1 iteration step, and the updated configuration will be closer to the equilibrium one. The out-of-balance forces of the system are defined as:

$$\boldsymbol{P}_a^i = \boldsymbol{E}_a^T(\boldsymbol{f}_{np} - \boldsymbol{g}) - \boldsymbol{K}_{aa}|_{\boldsymbol{n}^i}\boldsymbol{n}_a^i - \boldsymbol{K}_{ab}|_{\boldsymbol{n}^i}\boldsymbol{n}_b^i - \boldsymbol{K}_{Tab}(\boldsymbol{n}_b^{i+1} - \boldsymbol{n}_b^i). \tag{78}$$

The difference of the minimal coordinate can be simply computed by:

$$\mathrm{d}\boldsymbol{n}_a^i = \boldsymbol{K}_{Taa}^{-1}\boldsymbol{P}_a^i. \tag{79}$$

The above equation can solve nonlinear equilibrium equations based on an iteration method.

## 5.2. Linearized equilibrium equation in terms of the member force

The equilibrium equation Eq.(71) can be written linearly with the member force vector $\boldsymbol{t}_c$:

$$\boldsymbol{E}_a^T\boldsymbol{A}_{2c}\boldsymbol{t}_c = \boldsymbol{E}_a^T(\boldsymbol{f}_{np} - \boldsymbol{g}), \tag{80}$$

where $\mathbf{A}_{2c} \in \mathbb{R}^{3n_n \times n_{ec}}$ is the equilibrium matrix of clustered tensegrity considering pulley size:

$$\mathbf{A}_{2c} = \{(\boldsymbol{C}^T\hat{\boldsymbol{l}}_s\hat{\boldsymbol{l}}^{-2}\boldsymbol{C}) \otimes \boldsymbol{I}_3 - [\boldsymbol{C}^T(\widehat{\boldsymbol{C}\boldsymbol{\mu}\boldsymbol{R}})\hat{\boldsymbol{l}}^{-2}] \otimes \boldsymbol{I}_3\boldsymbol{I}_{ne} \otimes [\boldsymbol{z}]^\times\}\text{b.d.}(\boldsymbol{H})\boldsymbol{S}^T. \tag{81}$$

The derivation from Eq.(71) to Eq.(80) uses the following formulation and Eq.(16):

$$\begin{aligned}(\hat{\boldsymbol{x}}\boldsymbol{C}) \otimes \boldsymbol{I}_3\boldsymbol{n} &= \left(\boldsymbol{I}_{n_e} \otimes \boldsymbol{I}_{3,1}\boldsymbol{x}\right)^\wedge \boldsymbol{C} \otimes \boldsymbol{I}_3\boldsymbol{n} \\ &= (\boldsymbol{C} \otimes \boldsymbol{I}_3\boldsymbol{n})^\wedge \boldsymbol{I}_{n_e} \otimes \boldsymbol{I}_{3,1}\boldsymbol{x} \\ &= \boldsymbol{b.d.}(\boldsymbol{H})\boldsymbol{x}.\end{aligned} \tag{82}$$

Eq. (80) can be simplified to:

$$\bar{\mathbf{A}}_{2c}\boldsymbol{t}_c = \boldsymbol{E}_a^T\left(\boldsymbol{f}_{np} - \boldsymbol{g}\right), \tag{83}$$

where $\bar{\mathbf{A}}_{2c} \in \mathbb{R}^{n_a \times n_{ec}}$ is the equilibrium matrix considering boundary constraints and clustered members:

$$\bar{\mathbf{A}}_{2c} = \boldsymbol{E}_a^T\mathbf{A}_{2c}. \tag{84}$$

The singular value decomposition of equilibrium matrix $\bar{\mathbf{A}}_{2c}$ reveals the self-stress mode and mechanism mode of the structure [39]:

$$\bar{\mathbf{A}}_{2c} = \boldsymbol{W}\boldsymbol{\Sigma}\boldsymbol{V}^T = [\boldsymbol{W}_1 \quad \boldsymbol{W}_2]\begin{bmatrix}\boldsymbol{\Sigma}_0 & \boldsymbol{0} \\ \boldsymbol{0} & \boldsymbol{0}\end{bmatrix}\begin{bmatrix}\boldsymbol{V}_1^T \\ \boldsymbol{V}_2^T\end{bmatrix}, \tag{85}$$

where $\boldsymbol{W} \in \mathbb{R}^{n_a \times n_a}$, and $\boldsymbol{V} \in \mathbb{R}^{n_{ec} \times n_{ec}}$ are orthogonal matrices. Let $r = \text{rank}(\bar{\mathbf{A}}_{2c})$ be the rank of $\bar{\mathbf{A}}_{2c}$. $\boldsymbol{V}_1 \in \mathbb{R}^{n_{ec} \times \text{r}}$ and $\boldsymbol{V}_2 \in \mathbb{R}^{n_{ec} \times (n_{ec} - \text{r})}$ are the row space and null space of $\bar{\mathbf{A}}_{2c}$. $\boldsymbol{W}_1 \in \mathbb{R}^{n_a \times \text{r}}$ and $\boldsymbol{W}_2 \in \mathbb{R}^{n_a \times (n_a - \text{r})}$ are the column space and left null space of $\bar{\mathbf{A}}_{2c}$. Since $\bar{\mathbf{A}}_{2c}\boldsymbol{V}_2 = \boldsymbol{0}$ and $\bar{\mathbf{A}}_{2c}^{\text{T}}\boldsymbol{W}_2 = \boldsymbol{0}$, $\boldsymbol{V}_2$ and $\boldsymbol{W}_2$ are the self-stress mode and mechanism mode of the tensegrity structure.

## 5.3. Compatibility matrix

The compatibility equation is the relation between the nodal displacement and member length that checks whether the structural deformations are physically valid. Recall that for TTS [27], the compatibility equation is:

$$\boldsymbol{B}_l d\boldsymbol{n} = d\boldsymbol{l}, \tag{86}$$

where $\boldsymbol{B}_l \in \mathbb{R}^{n_e \times 3n_n}$ is the compatibility matrix for TTS with pure bar and string connection:

$$\boldsymbol{B}_l = \hat{\boldsymbol{l}}^{-1}b.d.(\boldsymbol{H})^T(\boldsymbol{C} \otimes \mathbf{I}_3). \tag{87}$$

For clustered tensegrity with clustered members, the compatibility equation is:

$$\boldsymbol{B}_{l_c} d\boldsymbol{n} = d\boldsymbol{l}_c, \tag{88}$$

where $\boldsymbol{B}_{lc} \in \mathbb{R}^{n_{ec} \times 3n_n}$ is the compatibility matrix of CTS. Take the derivative of Eq.(80). We have:

$$\boldsymbol{A}_{2c} d\boldsymbol{t}_c = d\boldsymbol{f}_{np}. \tag{89}$$

The principle of virtual work reveals that the work done by external force equals to the increasement of potential energy:

$$d\boldsymbol{f}_{np}^T d\boldsymbol{n} = d\boldsymbol{t}_c^T d\boldsymbol{l}_c. \tag{90}$$

Substitute Eq.(88) and Eq.(89) into Eq.(90), we have:

$$\boldsymbol{B}_{l_c} = \boldsymbol{A}_{2c}^T. \tag{91}$$

Taking minimal nodal coordinate as the variable, the compatibility equation is:

$$\overline{\boldsymbol{B}}_{lc} d\boldsymbol{n}_a = d\boldsymbol{l}_c. \tag{92}$$

The compatibility matrix can also be obtained by the principle of virtual work:

$$\overline{\boldsymbol{B}}_{lc} = \overline{\boldsymbol{A}}_{2c}^{T}. \tag{93}$$

## 5.4. Tangent stiffness matrix

The left half part of equilibrium equation Eq.(71) represents the nodal force generated by an inner force, which is a nonlinear function of $\boldsymbol{n}, \boldsymbol{l}_s, \boldsymbol{l},$ and $\boldsymbol{t}$. So, the tangent stiffness matrix can be written as:

$$\boldsymbol{K}_T = \frac{\partial^2 V_e}{\partial \boldsymbol{n}^2} = \frac{\boldsymbol{K}\partial(\boldsymbol{n})}{\partial \boldsymbol{n}^T} + \frac{\partial(\boldsymbol{K}\boldsymbol{n})}{\partial \boldsymbol{l}_s^T}\frac{\partial \boldsymbol{l}_s}{\partial \boldsymbol{n}^T} + \frac{\partial(\boldsymbol{K}\boldsymbol{n})}{\partial \boldsymbol{l}^T}\frac{\mathrm{d}\boldsymbol{l}}{\mathrm{d}\boldsymbol{n}^T} + \frac{\partial(\boldsymbol{K}\boldsymbol{n})}{\partial \boldsymbol{t}^T}\frac{\partial \boldsymbol{t}}{\partial \boldsymbol{n}^T}. \tag{94}$$

The first term of Eq.(94) is:

$$\frac{\boldsymbol{K}\partial \boldsymbol{n}}{\partial \boldsymbol{n}^T} = \boldsymbol{K}. \tag{95}$$

The second term of Eq.(94) can be obtained as:

$$\frac{\partial(\boldsymbol{K}\boldsymbol{n})}{\partial \boldsymbol{l}_s^T} = -\left(\boldsymbol{C}^T\hat{\boldsymbol{l}}^{-2}\hat{\boldsymbol{t}}\right) \otimes \boldsymbol{I}_3 \mathrm{b.d.}(\boldsymbol{H}), \tag{96}$$

and

$$\frac{\partial \boldsymbol{l}_S}{\partial \boldsymbol{n}^T} = \hat{\boldsymbol{l}}_S{}^{-1}\hat{\boldsymbol{l}}\boldsymbol{B}_l. \tag{97}$$

The third term of Eq.(94) is:

$$\frac{\partial(\boldsymbol{K}\boldsymbol{n})}{\partial \boldsymbol{l}^T} = -2\left(\boldsymbol{C}^T\hat{\boldsymbol{l}}_S\hat{\boldsymbol{t}}\right) \otimes \boldsymbol{I}_3 \mathrm{b.d.}(\boldsymbol{H})\hat{\boldsymbol{l}}^{-3} + 2\left[\boldsymbol{C}^T\left(\widehat{\boldsymbol{C}\boldsymbol{\mu}\boldsymbol{R}}\right)\hat{\boldsymbol{t}}\right] \otimes \boldsymbol{I}_3\boldsymbol{I}_{ne} \otimes [\boldsymbol{z}]^{\times}\mathrm{b.d.}(\boldsymbol{H})\hat{\boldsymbol{l}}^{-3}, \tag{98}$$

where:

$$\frac{\partial \boldsymbol{l}}{\partial \boldsymbol{n}^T} = \boldsymbol{B}_l. \tag{99}$$

The fourth term of Eq.(94) is:

$$\frac{\partial(\boldsymbol{K}\boldsymbol{n})}{\partial \boldsymbol{t}_c^T} = \boldsymbol{A}_{2c}, \tag{100}$$

and we have:

$$\frac{\partial \boldsymbol{t}_c}{\partial \boldsymbol{n}^T} = \widehat{\boldsymbol{E}}_c\widehat{\boldsymbol{A}}_c\hat{\boldsymbol{l}}_{0c}{}^{-1}\boldsymbol{A}_{2c}^T. \tag{101}$$

Substitute Eq.(95) to Eq.(101) into Eq.(94), and we have the tangent stiffness of clustered tensegrity considering pulley size:

$$\boldsymbol{K}_T = \left(\boldsymbol{C}^T\hat{\boldsymbol{l}}_S\hat{\boldsymbol{l}}^{-2}\hat{\boldsymbol{t}}\boldsymbol{C}\right) \otimes \boldsymbol{I}_3 - \left[\boldsymbol{C}^T\left(\widehat{\boldsymbol{C}\boldsymbol{\mu}\boldsymbol{R}}\right)\hat{\boldsymbol{l}}^{-2}\hat{\boldsymbol{t}}\right] \otimes \boldsymbol{I}_3\boldsymbol{I}_{ne} \otimes [\boldsymbol{z}]^{\times}\boldsymbol{C} \otimes \boldsymbol{I}_3 + \boldsymbol{B}_l^T\left[\hat{\boldsymbol{l}}^{-2}\hat{\boldsymbol{l}}_S{}^{-1}\left(\hat{\widehat{\boldsymbol{R}}}{}^2 - \hat{\boldsymbol{l}}_S{}^2\right)\hat{\boldsymbol{t}}\right]\boldsymbol{B}_l + 2\left[\boldsymbol{C}^T\left(\widehat{\boldsymbol{C}\boldsymbol{\mu}\boldsymbol{R}}\right)\hat{\boldsymbol{l}}^{-3}\hat{\boldsymbol{t}}\right] \otimes \boldsymbol{I}_3\boldsymbol{I}_{ne} \otimes [\boldsymbol{z}]^{\times}\mathrm{b.d.}(\boldsymbol{H})\boldsymbol{B}_l + \boldsymbol{A}_{2c}\widehat{\boldsymbol{E}}_c\widehat{\boldsymbol{A}}_c\hat{\boldsymbol{l}}_{0c}{}^{-1}\boldsymbol{A}_{2c}^T. \tag{102}$$

The tangent stiffness can be divided into two parts:

$$\boldsymbol{K}_T = \boldsymbol{K}_g + \boldsymbol{K}_e, \tag{103}$$

where $\boldsymbol{K}_{\mathrm{g}} \in \mathbb{R}^{3n_n \times 3n_n}$ is the geometry stiffness caused by the deformation of the structure while assuming the member force is constant. $\boldsymbol{K}_e$ is the material stiffness caused by the difference of member force caused by the elongation of members:

$$
\begin{aligned}
\boldsymbol{K}_g = \left(\boldsymbol{C}^T \hat{\boldsymbol{l}}_S \hat{\boldsymbol{l}}^{-2} \hat{\boldsymbol{t}} \boldsymbol{C}\right) \otimes \boldsymbol{I}_3 - \left[\boldsymbol{C}^T \left(\widehat{\boldsymbol{C}\hat{\boldsymbol{\mu}}\boldsymbol{R}}\right) \hat{\boldsymbol{l}}^{-2} \hat{\boldsymbol{t}}\right] \otimes \boldsymbol{I}_3 \boldsymbol{I}_{ne} \otimes [\boldsymbol{z}]^{\times} \boldsymbol{C} \otimes \boldsymbol{I}_3 + \\
\boldsymbol{B}_l^T \left[\hat{\boldsymbol{l}}^{-2} {\hat{\boldsymbol{l}}_S}^{-1} \left(\hat{\widehat{\boldsymbol{R}}}^2 - {\hat{\boldsymbol{l}}_S}^2\right) \hat{\boldsymbol{t}}\right] \boldsymbol{B}_l + 2\left[\boldsymbol{C}^T \left(\widehat{\boldsymbol{C}\hat{\boldsymbol{\mu}}\boldsymbol{R}}\right) \hat{\boldsymbol{l}}^{-3} \hat{\boldsymbol{t}}\right] \otimes \boldsymbol{I}_3 \boldsymbol{I}_{ne} \otimes [\boldsymbol{z}]^{\times} \mathrm{b.d.}(\boldsymbol{H}) \boldsymbol{B}_l,
\end{aligned}
\tag{104}
$$

$$
\boldsymbol{K}_e = \boldsymbol{A}_{2c} \widehat{\boldsymbol{E}}_c \widehat{\boldsymbol{A}}_c {\hat{\boldsymbol{l}}_{0c}}^{-1} \boldsymbol{A}_{2c}^T. \tag{105}
$$

Note that the first part of $\boldsymbol{K}_\mathrm{g}$ is identity to stiffness matrix. The material stiffness $\boldsymbol{K}_\mathrm{e}$ is consistent with that of clustered tensegrity without considering pulley size. Rewrite the tangent stiffness matrix of CTS [32]:

$$
\boldsymbol{K}_{T,CTS} = \left[\boldsymbol{C}^T \hat{\boldsymbol{l}}^{-1} (\widehat{\boldsymbol{S}^T \boldsymbol{t}_c}) \boldsymbol{C}\right] \otimes \boldsymbol{I}_3 - \boldsymbol{A}_2 \hat{\boldsymbol{l}}^{-1} \hat{\boldsymbol{t}} \boldsymbol{A}_2^T + \boldsymbol{A}_{2c} \widehat{\boldsymbol{E}}_c \widehat{\boldsymbol{A}}_c {\hat{\boldsymbol{l}}_{0c}}^{-1} \boldsymbol{A}_{2c}^T. \tag{106}
$$

We can observe that the tangent stiffness matrix of clustered tensegrity considering pulley size in Eq.(102) degenerates to clustered tensegrity not considering pulley size in Eq.(106) if $\boldsymbol{R}$=$\boldsymbol{0}$. Furthermore, the tangent stiffness matrix in Eq.(102) degenerates to TTS without considering clustered strings by setting $\boldsymbol{R}$=$\boldsymbol{0}$ and $\boldsymbol{S} = \boldsymbol{I}$ [40-43].

# 6. Numerical examples

In this section, three examples are given to study the influence of the pulley sizes on the structure's equilibrium configuration and member force, as well as demonstrate the accuracy and efficiency of the proposed equilibrium and stiffness of clustered tensegrity structures considering pulley sizes.

## 6.1. Compound pulley system

A compound pulley system example is shown in Figure 4, where the structure is composed of 1 clustered string (string 1-4), two individual strings (strings 5 and 6), one fixed pulley (at node 3), two free pulleys (at nodes 2 and 4), and two pinned nodes (at nodes 1 and 5). The clustered and individual strings are composed of six segments (strings 1-6, $n_e = 6$).

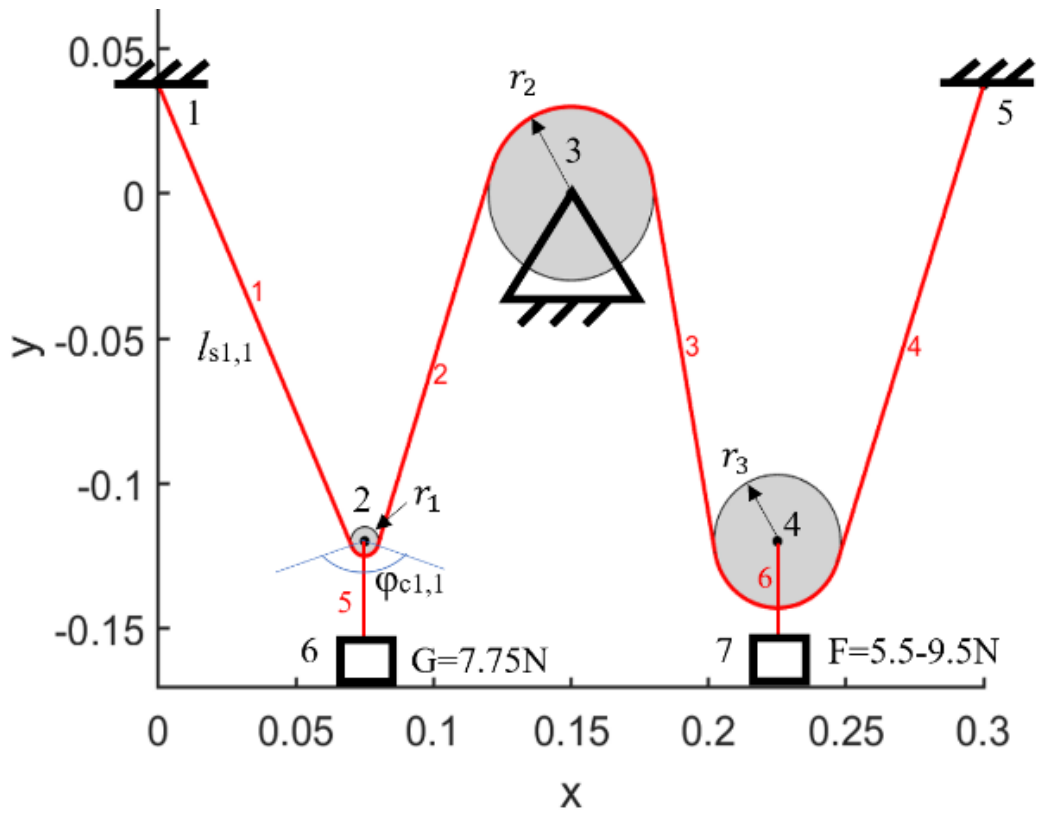

(a) $r_1$=5mm.

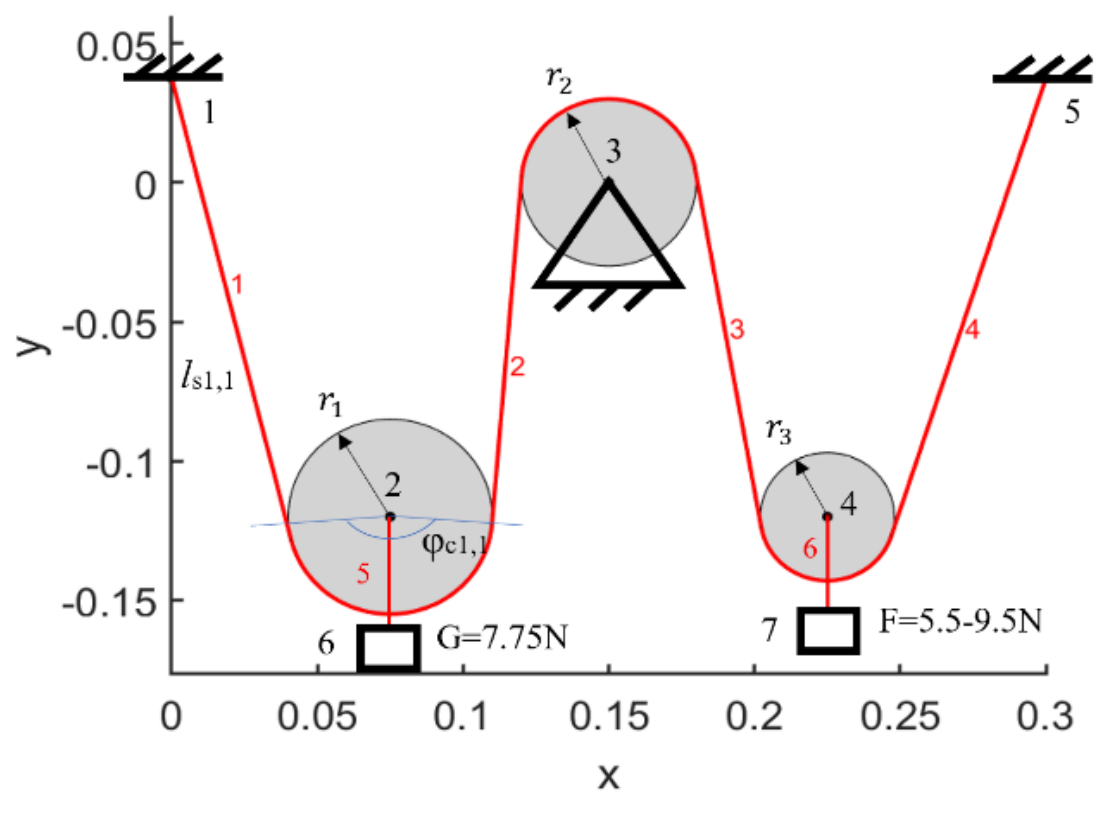

(b) $r_1$=35mm.

Figure 4 Configuration of the compound pulley system.

Since there are two strings connected to the free pulleys (at nodes 2 and 4), the nodal coordinate vector of the whole structure, according to Eq. (1), can be written as:

$$
n = [n_{1,1}^T, n_{2,1}^T, n_{2,2}^T, n_{3,1}^T, n_{4,1}^T, n_{4,2}^T, n_{5,1}^T, n_{6,1}^T, n_{7,1}^T]^T.
$$

The radius of the three pulleys is $r_1$, $r_2$, $r_3$. According to Eq. (5), the radius vector of the structure is:

$R = [R_{1,1}, R_{2,1}, R_{2,2}, R_{3,1}, R_{4,1}, R_{4,2}, R_{5,1}, R_{6,1}, R_{7,1}]^T = [0 \quad r_1 \quad 0 \quad r_2 \quad r_3 \quad 0 \quad 0 \quad 0 \quad 0]^T$.

The connectivity matrix of the structure is $C = \begin{bmatrix} -1 & 1 & 0 & 0 & 0 & 0 & 0 & 0 & 0 \\ 0 & -1 & 0 & 1 & 0 & 0 & 0 & 0 & 0 \\ 0 & 0 & 0 & -1 & 1 & 0 & 0 & 0 & 0 \\ 0 & 0 & 0 & 0 & -1 & 0 & 1 & 0 & 0 \\ 0 & 0 & -1 & 0 & 0 & 0 & 0 & 1 & 0 \\ 0 & 0 & 0 & 0 & 0 & -1 & 0 & 0 & 1 \end{bmatrix}$. The side vector of strings (the relative position of the string to the pulley) is:

$\mu = [\mu_{1,1}, \mu_{2,1}, \mu_{2,2}, \mu_{3,1}, \mu_{4,1}, \mu_{4,2}, \mu_{5,1}, \mu_{6,1}, \mu_{7,1}]^T = [0 \quad 1 \quad 0 \quad -1 \quad 1 \quad 0 \quad 0 \quad 0 \quad 0]^T$.

The clustered matrix is:

$$S = \begin{bmatrix} 1\,1\,1\,1\,0\,0 \\ 0\,0\,0\,0\,1\,0 \\ 0\,0\,0\,0\,0\,1 \end{bmatrix}.$$

A sequence matrix represents the relative position of segments in the clustered string:

$$S_{T1} = \begin{bmatrix} 1\,0\,0\,0\,0\,0 \\ 0\,1\,0\,0\,0\,0 \\ 0\,0\,1\,0\,0\,0 \\ 0\,0\,0\,1\,0\,0 \end{bmatrix}, \; S_{T2} = [0\,0\,0\,0\,1\,0], \; S_{T3} = [0\,0\,0\,0\,0\,1].$$

The Young's modulus of all the strings is set to be $2 \times 10^{11}$ Pa and the cross-sectional area of the strings are $2.5 \times 10^{-6} \text{m}^2$. The radius of the fixed pulley is $r_2 = 30\text{mm}$, and the radius of the free pulley in the right is $r_3 = 23\text{mm}$. The radius of the left pulley is changed from $r_1 = 5\text{mm}$ to $r_1 = 35\text{mm}$ with an increment of 10mm. The constant load of 7.75N is applied in the negative Y-direction on the $n_6$, and a load from 5.5N to 9.5N is equally applied within 5 substeps in the negative Y-direction on the $n_7$. The equilibrium configuration is obtained by solving the nonlinear equilibrium equation of Eq. (71), and the results in each substep are shown in Figures 5 to 7. The length of the straight part of the first segment in the clustered string $l_{s1,1}$ with different pulley sizes is shown in Figure 5. It can be observed that the length of the straight part of the first segment in the clustered string decreases as the radius of the first pulley increases. The Y-coordinate of node 2 with different pulley sizes is shown in Figure 6. And we can see that the nodal coordinate increases as the radius increases. As shown in Figure 7, the angle of the circular string in node 2 with a bigger pulley size generally has a larger angle. Bus as the external load increases, the angle gradually tends to be the same. As $r$ increases 10mm, the straight length decreases by about 12.5mm, and the Y-coordinates of node 2 increase about 12.5mm.

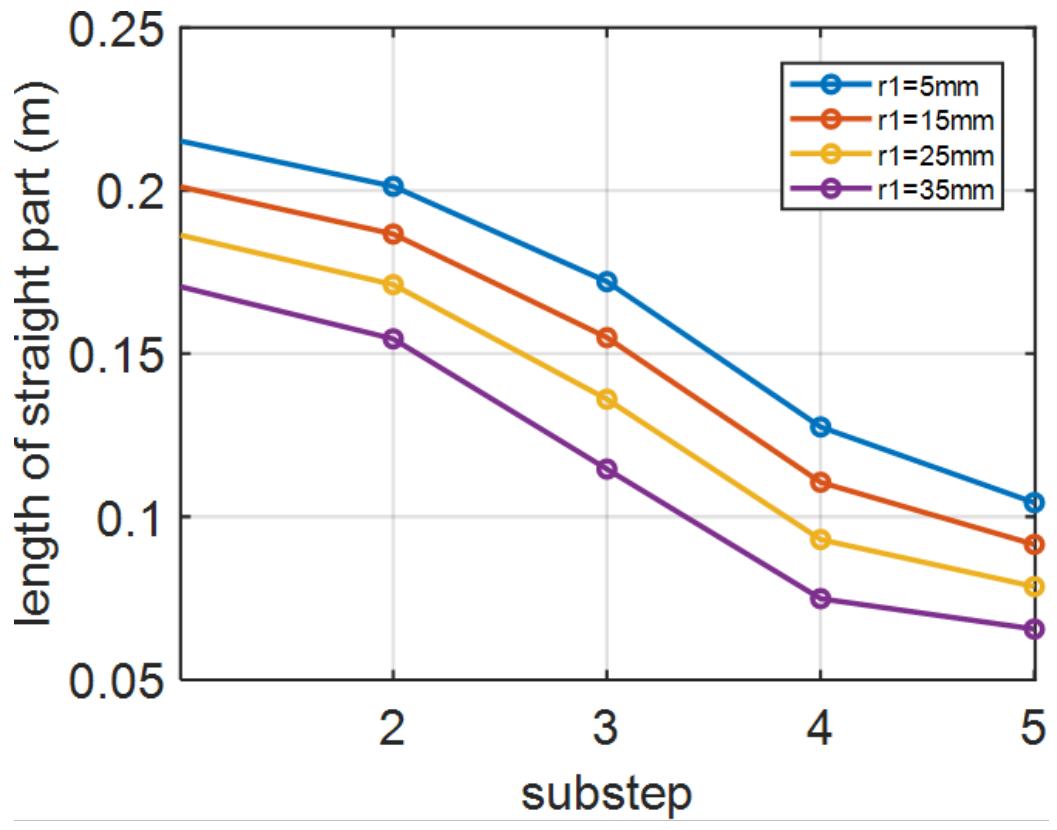

Figure 5 Length of the straight part $l_{s1,1}$ of the clustered strings.

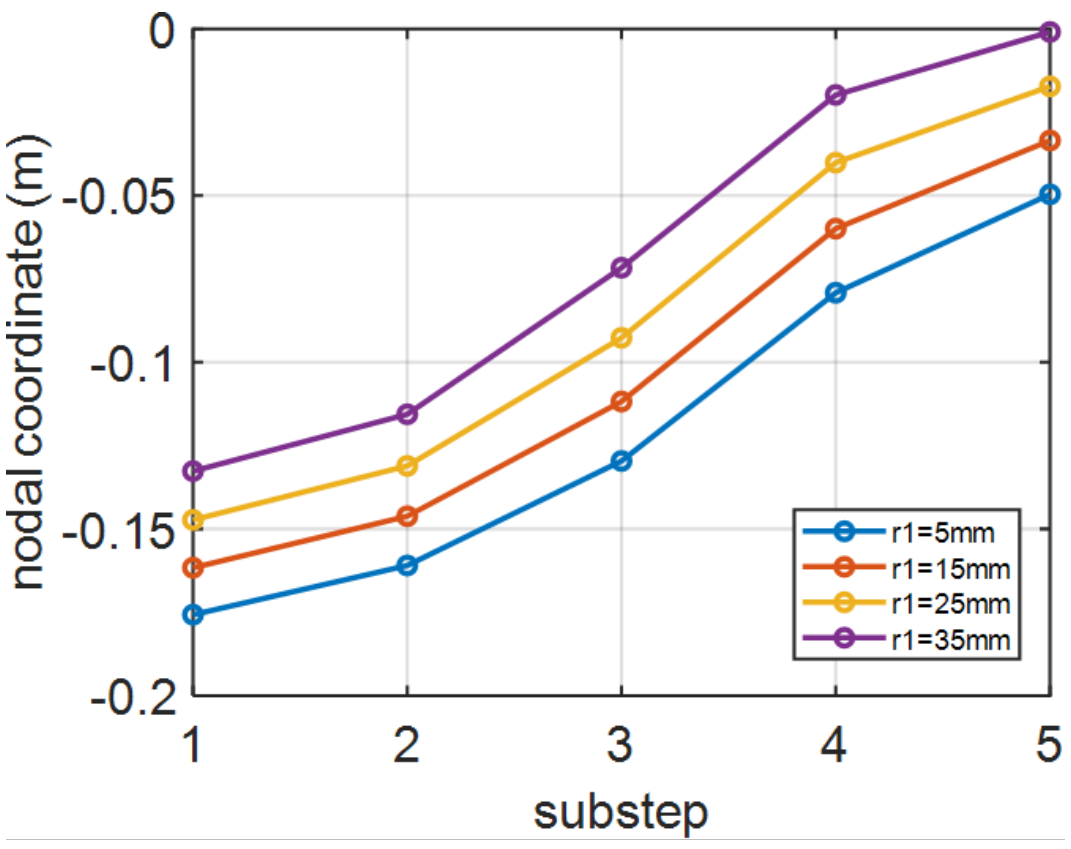

Figure 6 The $Y$-coordinate of node 2 with different sizes of pulley 1.

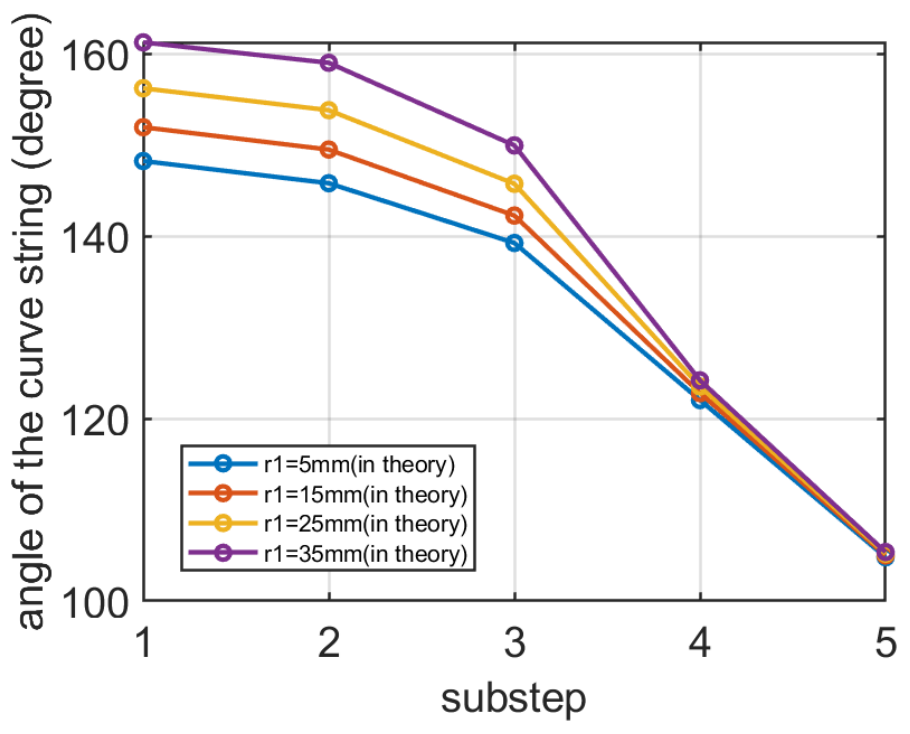

Figure 7 $\varphi_{c1,1}$ at each substep in the equilibrium simulation.

## 6.2. Clustered T-bar

In this section, we take the clustered T-bar as an example to study the influence of pulley size on the configuration and equilibrium of the CTS. A classical T-bar structure comprises four separate strings and two bars [32]. We transform it into a cluster one by connecting the members 3 and 4 into one clustered string, as shown in Figure 8. Note that bars are in black, and strings are in red. We set Young's modulus of members to be $2 \times 10^{11}$Pa. The cross-sectional area of bars and strings are $3.6 \times 10^{-5}\mathrm{m}^2$ and $4 \times 10^{-6}\mathrm{m}^2$. The rest length of bars, clustered string, and traditional string are respectively 2m, 2.8m, and 1.4m.

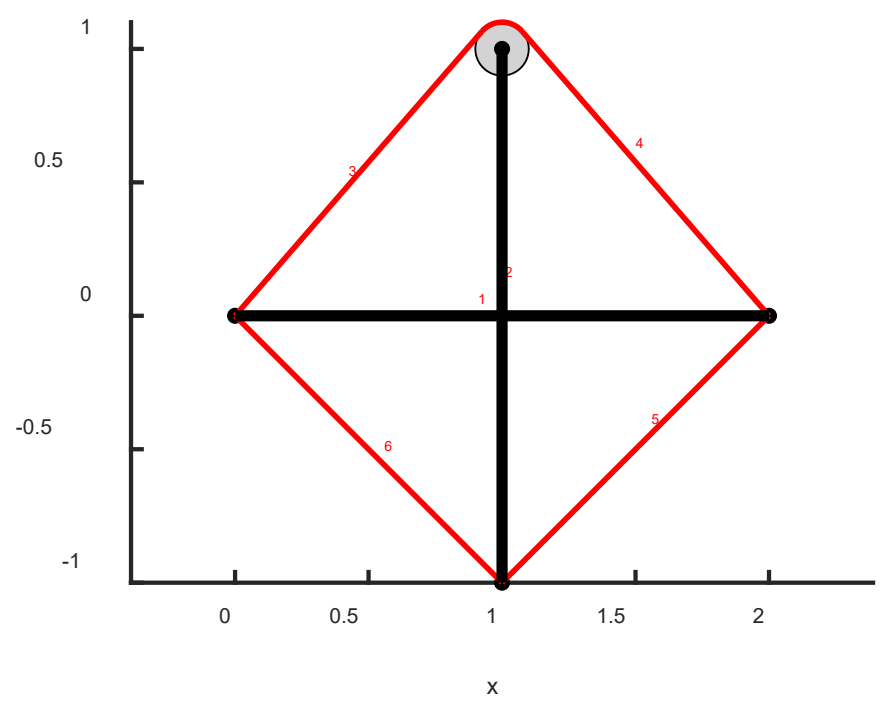

Figure 8 Configuration of the clustered T-bar structure.

We vary the pulley diameters to study pulley sizes' influence on the structure's equilibrium configuration, prestress, and stiffness. The initial configuration is shown in Figure 8. The equilibrium results show that as the diameter of the pulley increases, the clustered string's length increases, as shown in Figure 9, and the force of members increases, as in Figure 10. It also can be observed that the prestress of the structure is very sensitive to the pulley size. A 1mm increment in the pulley diameters will result in a $10^3$N in the structure member force. This is not a trivial phenomenon because a 0.1% increase in the strain of a steel string caused by the error of pulley sizes may lead to structure member failure in civil engineering practice.

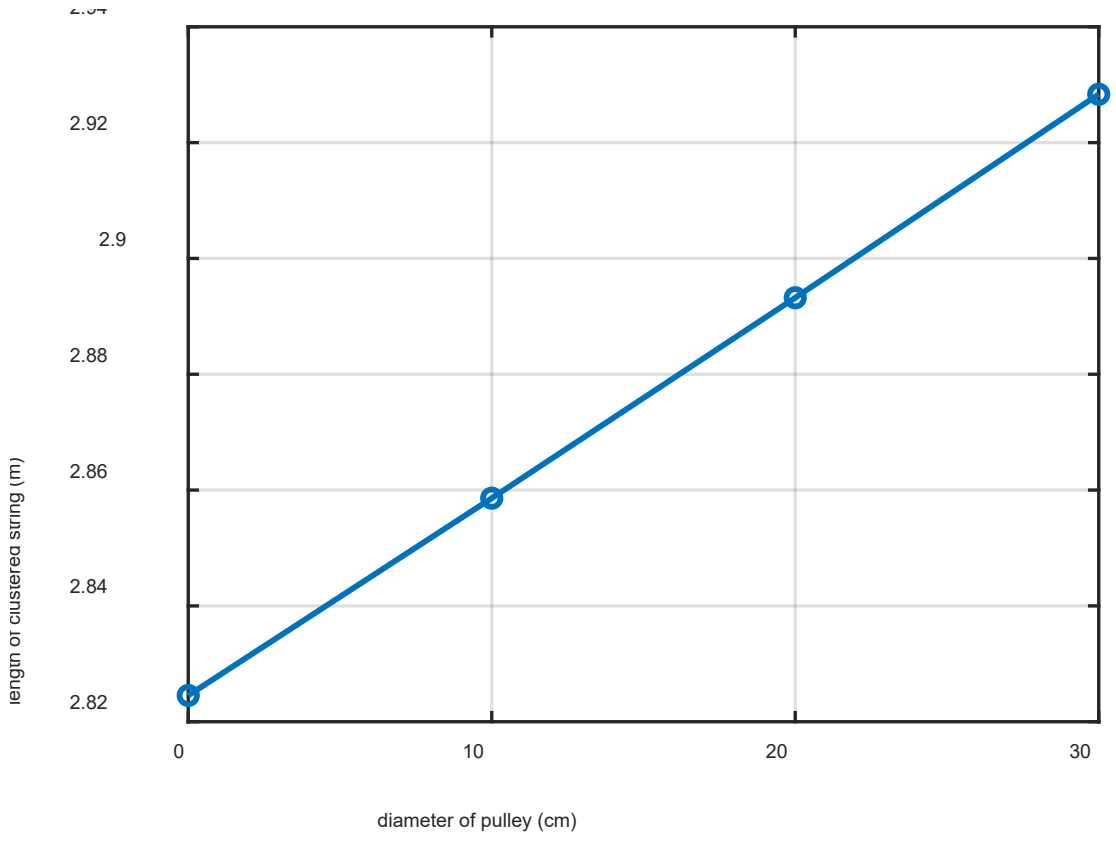

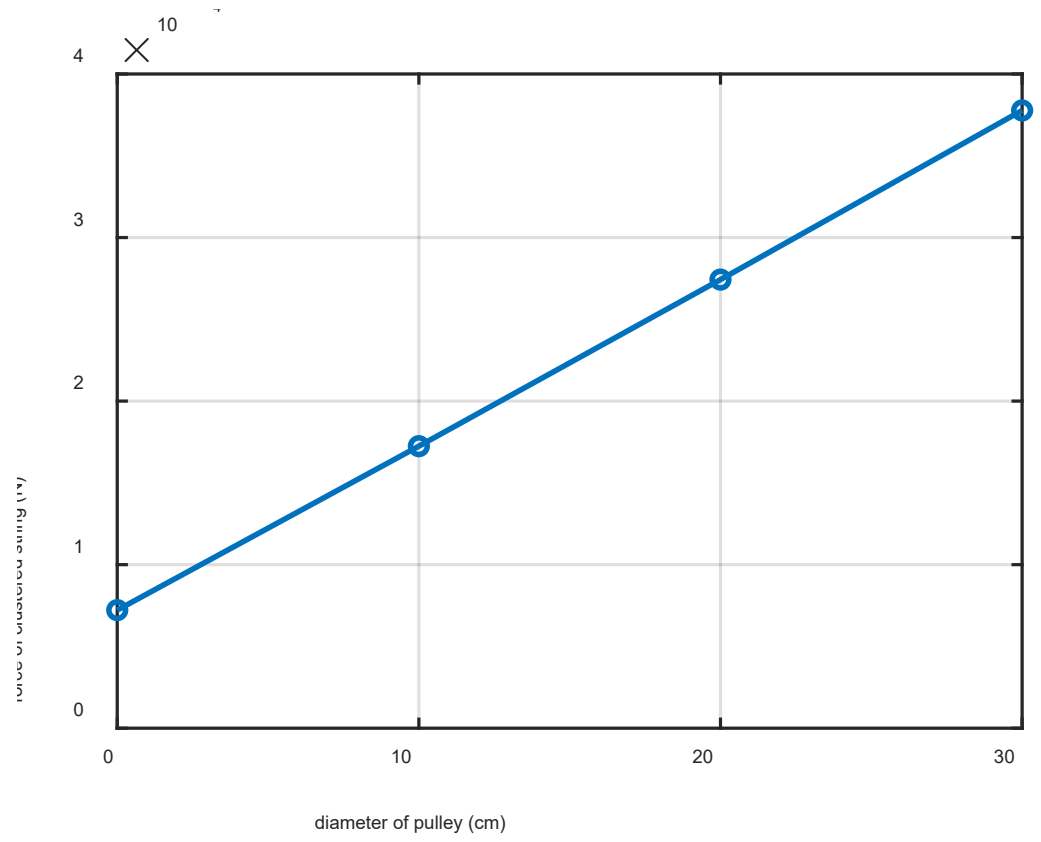

Figure 9 The length of clustered strings in different pulley sizes.

Figure 10 The force of clustered strings in different pulley sizes.

The clustered T-bar structure in different diameters has different structural stiffness, as shown in Figure 11. The minimal eigenvalue of the tangent stiffness matrix increases as the diameter of the pulley increases. This is because the prestress helps increase the structure's geometry stiffness. The mode shape of the tangent stiffness of the T-bar with a 30cm diameter pulley is shown in Figure 12. The first mode corresponds to the movement of the pulley node, and the first mode is also the most compliant mode, with about 0.35% stiffness of the second one.

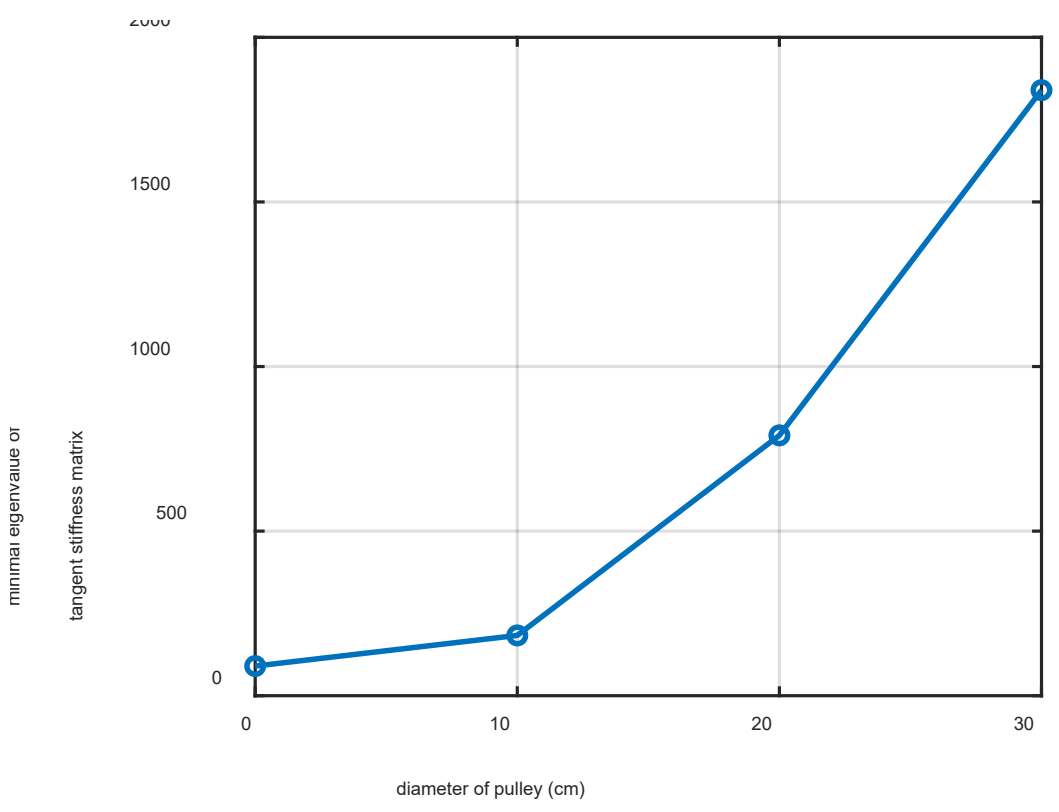

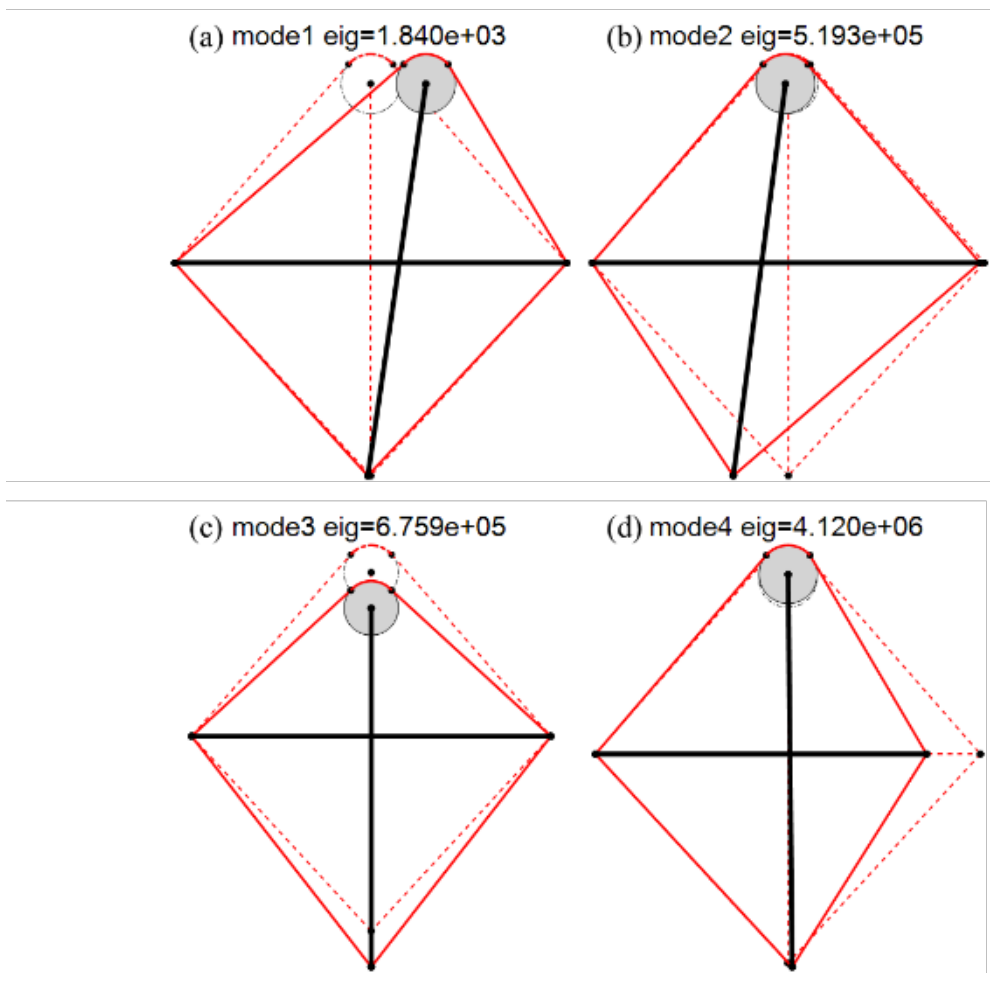

Figure 11 The minimal eigenvalue of tangent stiffness of T-bar with different pulley sizes.

Figure 12 The mode shapes from the tangent stiffness matrix.

## 6.3. Under-actuated robotic finger

In this section, an under-actuated robotic finger composed of tendon-pulley transmission and double-stage mechanisms, as shown in Figure 13, is studied by the proposed statics formulation of clustered tensegrity considering pulley sizes. The robotic finger is composed of 9 bars, 2 cables, 8 pulleys, and 2 pinned nodes. One end of the cable is tied to the top node of the diagonal line, and the other end of the cable is winded on the roller pinned to the ground, which is driven by a motor. The robotic finger bend to the left is, cable 1 is shortened, and cable 2 is elongated. The finger which owns more than 2-DOF is actuated by only one motor in the process of a closing finger. Thus, the finger belongs to a class of under-actuated mechanisms.

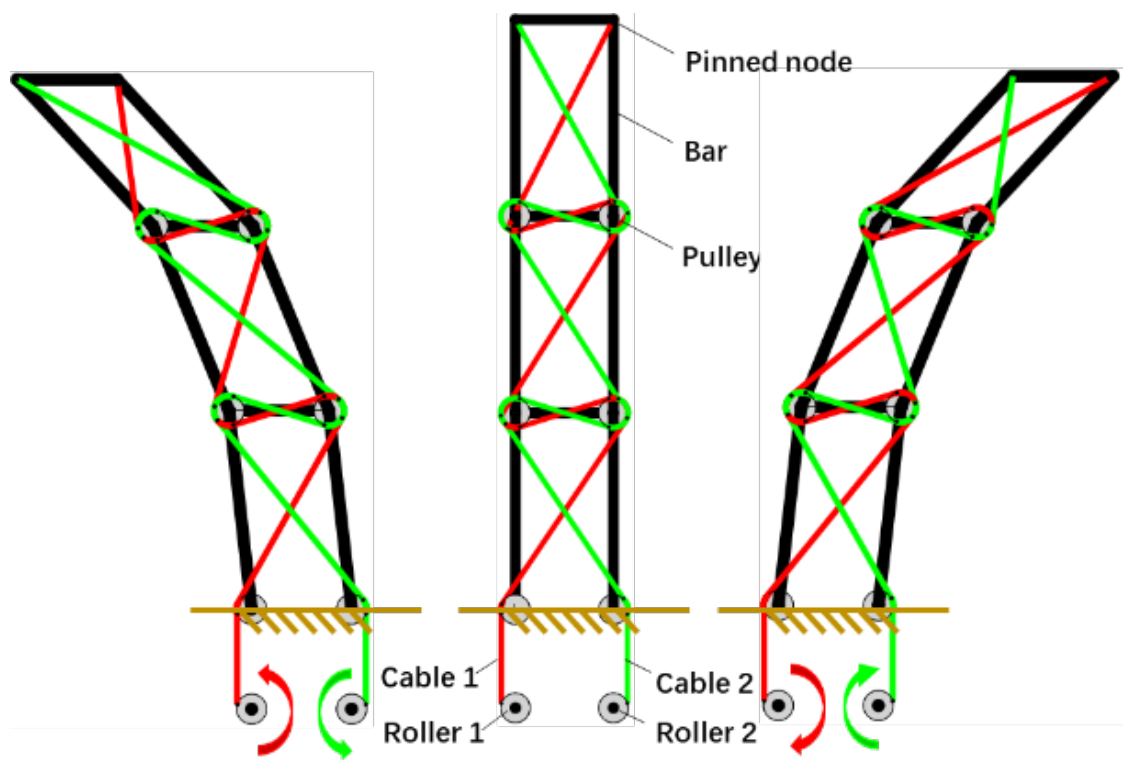

Figure 13 Schematic diagram of the under-actuated robotic finger.

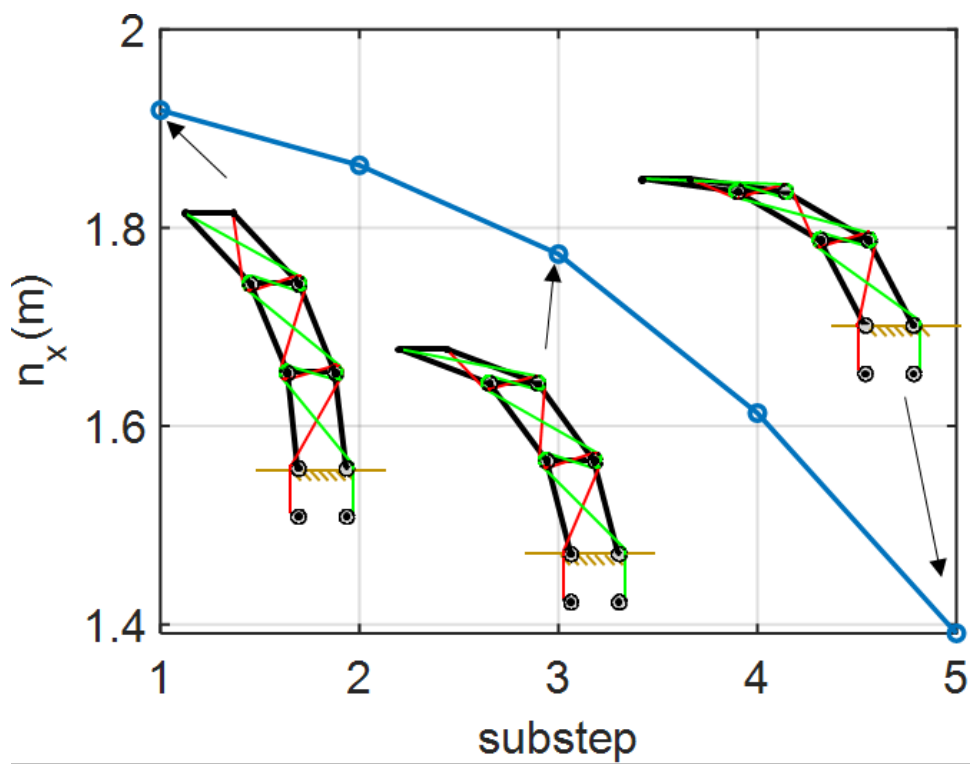

Figure 14 The X-coordinate of the top node to substeps.

The different radii of the pulleys are considered in the statics simulation. As shown in Figure 14, the robotic finger is composed of pulleys with a radius of 0.075 m. The rest of the length of cable 1 is changed from 4.9 m to 3.9 m equally in five substeps, while cable 2 is relaxed with no tension. The finger gradually bends to the left at each substep, and the top node's X coordinate and the structure's equilibrium configuration are shown in Figure 14. We evenly increase the pulley radius from 0m to 0.3m and decrease the rest length of cable 1 from 4.9m to 3.9 m at the 5 substeps. The X-coordinates of the top-left node vs. pulley sizes are shown in Figure 15. The length of cable 2 in the equilibrium configuration is shown in Figure 16. We can observe that the pulley radius will influence the equilibrium configuration and rest length of the under-actuated robotic finger cables.

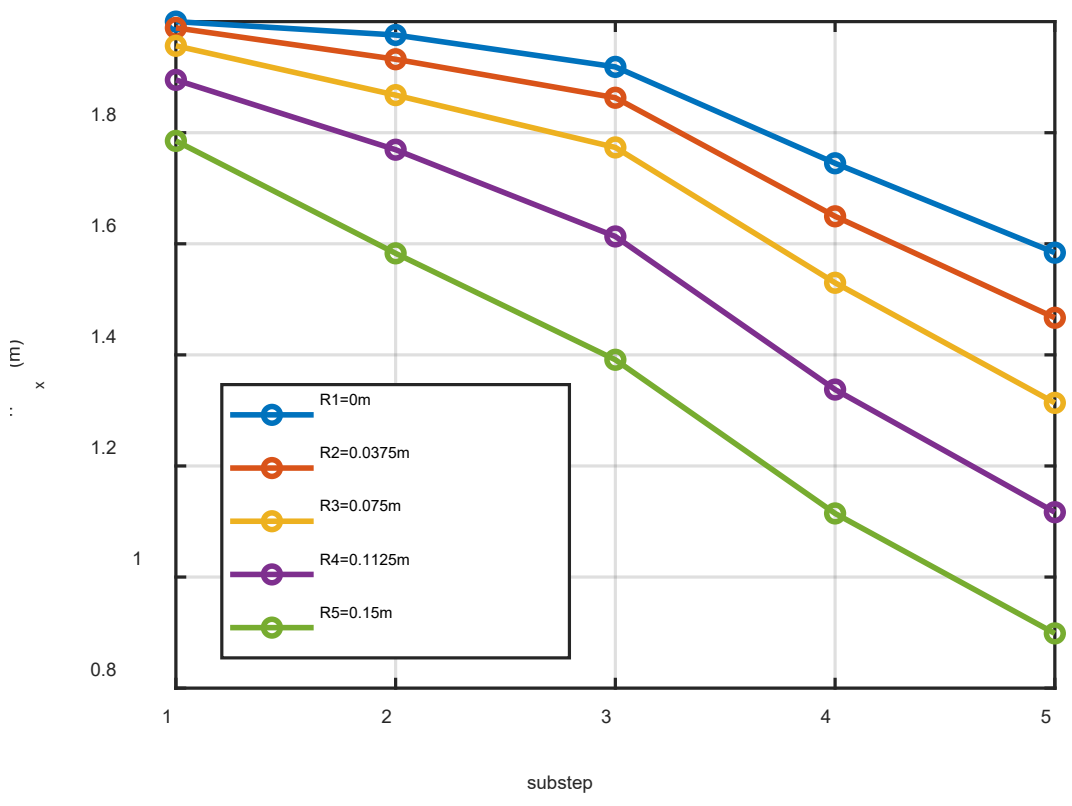

Figure 15 The $X$-coordinate of the top left node in different pulley sizes.

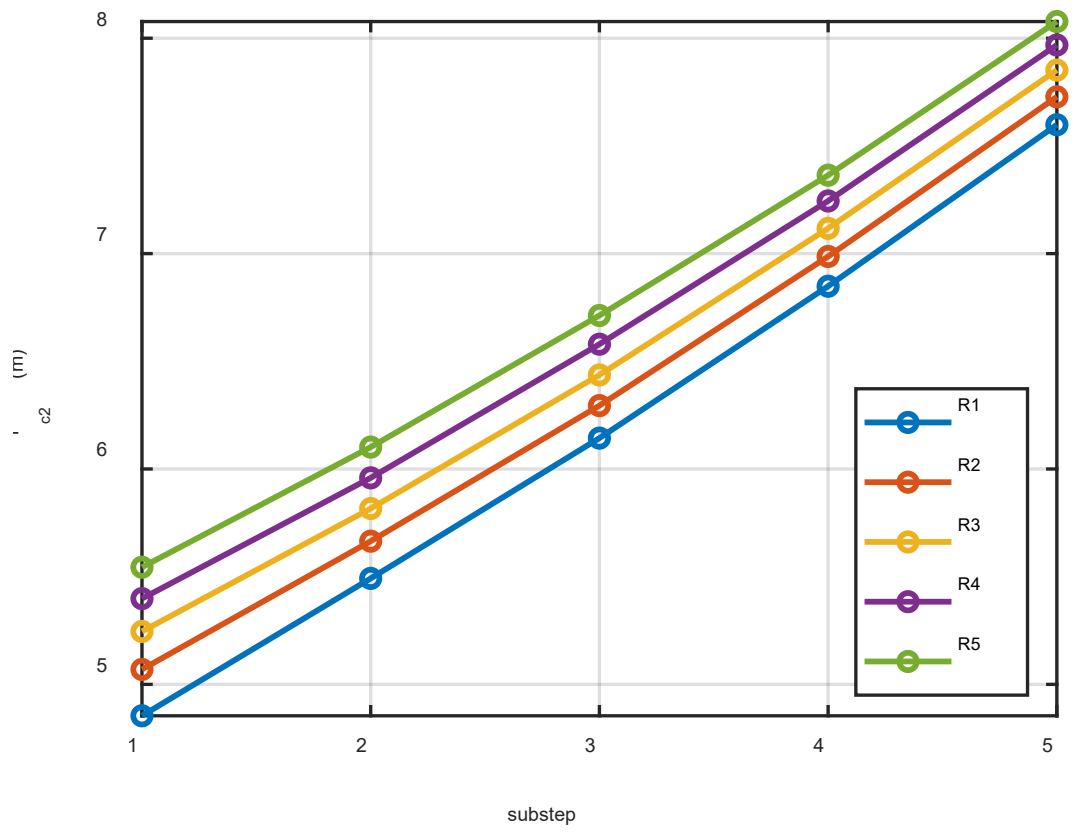

Figure 16 The length of the second clustered string (The green one on the right).

## 7. Conclusions

This paper presents a general approach to CTS's equilibrium and stiffness study considering pulley sizes. The equilibrium equations are derived based on the Lagrangian method, and the structure stiffness equations are given in detail. The proposed method allows one to conduct a quantitative analysis of the influence of pulleys on the clustered tensegrity. Results show that as pulley sizes go to zero, the governing equations of the CTS with pulleys yield to the classical CTS without pulleys in the existing literature. It is also shown that the prestress in the CTS is very sensitive to the size of the pulleys. For example, the clustered T-bar structure shows that an increment of 1mm in the pulley size can result in the force increment to 1e3N, which may lead to string failures. In other words, pulleys must be considered when putting CTS into engineering practice. This research can analyze a wide range of engineering structures with pulleys involved, such as compound pulley systems, clustered tensegrity systems, and robotics with clustering strategies.

## Acknowledgment

The research was supported by the Foundation of Key Laboratory of Space Structures of Zhejiang Province (Grant No. 202102) and the National Natural Science Foundation of China (Grant No. 51878600).

## Appendix A

The roof of Eq.(46) is given as follows:

$$
\begin{aligned}
\frac{\partial \varphi_{R_{i,j}}}{\partial n} &= \frac{\eta_{ij}}{\sin\varphi_{R_{i,j}}}\left[\frac{\boldsymbol{C}_{i,j}^T \otimes \boldsymbol{I}_3 \boldsymbol{h}_{i,j+1} + \boldsymbol{C}_{i,j+1}^T \otimes \boldsymbol{I}_3 \boldsymbol{h}_{i,j}}{|\boldsymbol{h}_{i,j}||\boldsymbol{h}_{i,j+1}|} - \frac{\boldsymbol{h}_{i,j}^T \boldsymbol{h}_{i,j+1}(\boldsymbol{C}_{i,j+1}^T \boldsymbol{C}_{i,j+1}) \otimes \boldsymbol{I}_3 \boldsymbol{n}}{|\boldsymbol{h}_{i,j}||\boldsymbol{h}_{i,j+1}|^3} - \frac{\boldsymbol{h}_{i,j}^T \boldsymbol{h}_{i,j+1}(\boldsymbol{C}_{i,j}^T \boldsymbol{C}_{i,j}) \otimes \boldsymbol{I}_3 \boldsymbol{n}}{|h_{i,j}|^3 |h_{i,j+1}|}\right] \\
&= \frac{\eta_{ij}}{\sin\varphi_{R_{i,j}} |\boldsymbol{h}_{i,j}||\boldsymbol{h}_{i,j+1}|}\left\{\frac{\boldsymbol{C}_{i,j}^T \otimes \boldsymbol{I}_3}{|\boldsymbol{h}_{i,j}|^2}\left[|\boldsymbol{h}_{i,j}|^2 \boldsymbol{h}_{i,j+1} - \boldsymbol{h}_{ij}(\boldsymbol{h}_{ij}^T \boldsymbol{h}_{i,j+1})\right] + \frac{\boldsymbol{C}_{i,j+1}^T \otimes \boldsymbol{I}_3}{|\boldsymbol{h}_{i,j+1}|^2}\left[|\boldsymbol{h}_{i,j+1}|^2 \boldsymbol{h}_{i,j} - \boldsymbol{h}_{i,j+1}(\boldsymbol{h}_{ij}^T \boldsymbol{h}_{i,j+1})\right]\right\} \\
&= \frac{\eta_{ij}}{\sin\varphi_{R_{i,j}} |\boldsymbol{h}_{i,j}||\boldsymbol{h}_{i,j+1}|}\left[\frac{\boldsymbol{C}_{i,j}^T \otimes \boldsymbol{I}_3}{|\boldsymbol{h}_{i,j}|^2} \boldsymbol{h}_{i,j} \times (\boldsymbol{h}_{i,j+1} \times \boldsymbol{h}_{i,j}) + \frac{\boldsymbol{C}_{i,j+1}^T \otimes \boldsymbol{I}_3}{|\boldsymbol{h}_{i,j+1}|^2} \boldsymbol{h}_{i,j+1} \times (\boldsymbol{h}_{i,j} \times \boldsymbol{h}_{i,j+1})\right] \\
&= \frac{\eta_{ij}}{\sin\varphi_{R_{i,j}} |\boldsymbol{h}_{i,j}||\boldsymbol{h}_{i,j+1}|}\underline{\mu}_{ij}\eta_{ij}\left[\frac{\boldsymbol{C}_{i,j}^T \otimes \boldsymbol{I}_3}{|\boldsymbol{h}_{i,j}|^2} \boldsymbol{h}_{i,j} \times \left(-\sin\varphi_{R_{i,j}} |\boldsymbol{h}_{i,j}||\boldsymbol{h}_{i,j+1}|\boldsymbol{z}\right)\right. \\
&\qquad \left. + \frac{\boldsymbol{C}_{i,j+1}^T \otimes \boldsymbol{I}_3}{|\boldsymbol{h}_{i,j+1}|^2} \boldsymbol{h}_{i,j+1} \times \left(\sin\varphi_{R_{i,j}} |\boldsymbol{h}_{i,j}||\boldsymbol{h}_{i,j+1}|\boldsymbol{z}\right)\right] \\
&= \underline{\mu}_{i,j}\left[\frac{\boldsymbol{C}_{i,j}^T \otimes \boldsymbol{I}_3}{|\boldsymbol{h}_{i,j}|^2} \boldsymbol{z} \times \boldsymbol{h}_{i,j} - \frac{\boldsymbol{C}_{i,j+1}^T \otimes \boldsymbol{I}_3}{|\boldsymbol{h}_{i,j+1}|^2} \boldsymbol{z} \times \boldsymbol{h}_{i,j+1}\right] \\
&= \underline{\mu}_{i,j}\left[\left(\boldsymbol{S}_{T_{i,j}} \boldsymbol{l}\right)^{-2} \left(\boldsymbol{S}_{T_{i,j}} \boldsymbol{C}\right)^T \otimes \boldsymbol{I}_3 \boldsymbol{z} \times \left(\boldsymbol{S}_{T_{i,j}} \boldsymbol{C}\right) \otimes \boldsymbol{I}_3 \boldsymbol{n} - \left(\boldsymbol{S}_{T_{i,j+1}} \boldsymbol{l}\right)^{-2} \left(\boldsymbol{S}_{T_{i,j+1}} \boldsymbol{C}\right)^T \otimes \boldsymbol{I}_3 \boldsymbol{z} \times \left(\boldsymbol{S}_{T_{i,j+1}} \boldsymbol{C}\right) \otimes \boldsymbol{I}_3 \boldsymbol{n}\right]
\end{aligned}
\tag{107}
$$

where $\boldsymbol{a} \times (\boldsymbol{b} \times \boldsymbol{c}) = (\boldsymbol{a}^T \boldsymbol{c})\boldsymbol{b} - (\boldsymbol{a}^T \boldsymbol{b})\boldsymbol{c}$ is used in the derivation.

# References:

[1] Pizzigoni A, Micheletti A, Ruscica G, Paris V, Bertino S, Mariani M, Trianni V, Madaschi S, A new T4 configuration for a deployable tensegrity pavilion, International Association for Shell and Spatial Structures (IASS), 2019, pp. 1-6.

[2] Bolelli PM, Rocha KB, De Oliveira IM, Mello WB, Haydamus AH, Boechat LC, Pauletti RM, Arcaro VF, Design and construction of a membrane-tensegrity sculpture, International Association for Shell and Spatial Structures (IASS), 2018, pp. 1-8.

[3] Fraddosio A, Pavone G, Piccioni MD. Minimal mass and self-stress analysis for innovative V-Expander tensegrity cells. Compos Struct. 2019;209: 754-774.

[4] Santos F, Caroço C, Amendola A, Miniaci M, Fraternali F. 3D tensegrity braces with superelastic response for seismic control. Int J Multiscale Com. 2022.

[5] Osikowicz NS, Roffman KM, Singla P, Lesieutre GA, Experimental shape control of cylindrical

triplex tensegrity structures, SPIE, 2022, pp. 129-143.
[6] Luo A, Liu H. Analysis for feasibility of the method for bars driving the ball tensegrity robot. Journal of Mechanisms and Robotics. 2017;9(5): 51010.
[7] Shah DS, Booth JW, Baines RL, Wang K, Vespignani M, Bekris K, Kramer-Bottiglio R. Tensegrity robotics. Soft Robot. 2021.
[8] Lee H, Jang Y, Choe JK, Lee S, Song H, Lee JP, Lone N, Kim J. 3D-printed programmable tensegrity for soft robotics. Science Robotics. 2020;5(45): y9024.
[9] Sabelhaus AP, Li AH, Sover KA, Madden JR, Barkan AR, Agogino AK, Agogino AM. Inverse statics optimization for compound tensegrity robots. IEEE Robotics and Automation Letters. 2020;5(3): 3982-3989.
[10] Fraternali F, Carpentieri G, Amendola A, Skelton RE, Nesterenko VF. Multiscale tunability of solitary wave dynamics in tensegrity metamaterials. Appl Phys Lett. 2014;105(20): 201903.
[11] Yin X, Gao Z, Zhang S, Zhang L, Xu G. Truncated regular octahedral tensegrity-based mechanical metamaterial with tunable and programmable Poisson's ratio. Int J Mech Sci. 2020;167: 105285.
[12] Bauer J, Kraus JA, Crook C, Rimoli JJ, Valdevit L. Tensegrity metamaterials: Toward failure‐resistant engineering systems through delocalized deformation. Adv Mater. 2021;33(10): 2005647.
[13] Fraddosio A, Pavone G, Piccioni MD. A novel method for determining the feasible integral self-stress states for tensegrity structures. Curved and Layered Structures. 2021;8(1): 70-88.
[14] Zhu D, Deng H, Wu X. Selecting active members to drive the mechanism displacement of tensegrities. Int J Solids Struct. 2020;191-192: 278-292.
[15] Lee S, Gan BS, Lee J. A fully automatic group selection for form-finding process of truncated tetrahedral tensegrity structures via a double-loop genetic algorithm. Composites Part B: Engineering. 2016;106: 308-315.
[16] Roth JK, McCarthy TJ. Optimizing compressive load capacity for differing tensegrity geometries. Comput Struct. 2021;249: 106523.
[17] Wang Y, Xu X, Luo Y. A unifying framework for form-finding and topology-finding of tensegrity structures. Comput Struct. 2021;247: 106486.
[18] Nie R, He B, Hodges DH, Ma X. Form finding and design optimization of cable network structures with flexible frames. Comput Struct. 2019;220: 81-91.
[19] Dong W, Stafford PJ, Ruiz-Teran AM. Inverse form-finding for tensegrity structures. Comput Struct. 2019;215: 27-42.
[20] Moored KW, Bart-Smith H. Investigation of clustered actuation in tensegrity structures. Int J Solids Struct. 2009;46(17): 3272-3281.
[21] You Z, Pellegrino S. Cable-stiffened pantographic deployable structures part 2: Mesh reflector. Aiaa J. 1997;35(8): 1348-1355.
[22] You Z, Pellegrino S. Cable-stiffened pantographic deployable structures. I-Triangular mast. Aiaa J. 1996;34(4): 813-820.
[23] Rhode-Barbarigos L, Ali NBH, Motro R, Smith IF. Design aspects of a deployable tensegrity-hollow-rope footbridge. International Journal of Space Structures. 2012;27(2-3): 81-95.
[24] Gomez-Jauregui V, Quilligan M, Manchado C, Otero C. Design, fabrication and construction of a deployable double-layer tensegrity grid. Struct Eng Int. 2018;28(1): 13-20.
[25] Chen ZH, Wu YJ, Yin Y, Shan C. Formulation and application of multi-node sliding cable element for the analysis of Suspen-Dome structures. Finite Elem Anal Des. 2010;46(9): 743-750.
[26] Ma S, Chen M, Yuan X, Skelton RE. Design and analysis of deployable clustered tensegrity cable

domes. arXiv preprint arXiv:2106.08424. 2021.
[27] Ma S, Chen M, Skelton RE. Tensegrity system dynamics based on finite element method. Compos Struct. 2022;280: 114838.
[28] Bel Hadj Ali N, Rhode-Barbarigos L, Smith IFC. Analysis of clustered tensegrity structures using a modified dynamic relaxation algorithm. Int J Solids Struct. 2011;48(5): 637-647.
[29] Zhang L, Gao Q, Liu Y, Zhang H. An efficient finite element formulation for nonlinear analysis of clustered tensegrity. Eng Computation. 2016;33(1): 252-273.
[30] Zhang L, Lu MK, Zhang HW, Yan B. Geometrically nonlinear elasto-plastic analysis of clustered tensegrity based on the co-rotational approach. Int J Mech Sci. 2015;93: 154-165.
[31] Kan Z, Peng H, Chen B, Zhong W. Nonlinear dynamic and deployment analysis of clustered tensegrity structures using a positional formulation FEM. Compos Struct. 2018;187.
[32] Ma S, Chen M, Skelton RE. Dynamics and control of clustered tensegrity systems. Eng Struct. 2022;264: 114391.
[33] Feng X, Ou Y, Miah MS. Energy‐based comparative analysis of optimal active control schemes for clustered tensegrity structures. Structural Control and Health Monitoring. 2018;25(10): e2215.
[34] Bel Hadj Ali N, Aloui O, Rhode-Barbarigos L. A finite element formulation for clustered cables with sliding-induced friction. International journal of space structures. 2022;37(2): 81-93.
[35] Miyasaka M, Haghighipanah M, Li Y, Matheson J, Lewis A, Hannaford B. Modeling Cable-Driven robot with hysteresis and cable–pulley network friction. IEEE/ASME Transactions on Mechatronics. 2020;25(2): 1095-1104.
[36] Zhang Z, Xie G, Shao Z, Gosselin C. Kinematic calibration of Cable-Driven parallel robots considering the pulley kinematics. Mech Mach Theory. 2022;169: 104648.
[37] Xue R, Du Z, Yan Z, Ren B. An estimation method of grasping force for laparoscope surgical robot based on the model of a cable-pulley system. Mech Mach Theory. 2019;134: 440-454.
[38] Liu K, Paulino GH. Nonlinear mechanics of non-rigid origami: An efficient computational approach. Proceedings of the Royal Society A: Mathematical, Physical and Engineering Sciences. 2017;473(2206): 20170348.
[39] Pellegrino S. Structural computations with the singular value decomposition of the equilibrium matrix. International Journal of Solids & Structures. 1993;30(21): 3025-3035.
[40] Yuan X, Ma S, Jiang S. Form-finding of tensegrity structures based on the Levenberg–Marquardt method. Computers and Structures. 2017;192: 171-180.
[41] Nagase K, Skelton RE. Minimal mass tensegrity structures. Journal of the International Association for Shell and Spatial Structures. 2014;55(1): 37-48.
[42] Guest SD. The stiffness of tensegrity structures. Ima J Appl Math. 2011;76(1): 57-66.
[43] Zhang JY, Ohsaki M. Stability conditions for tensegrity structures. International Journal of Solids & Structures. 2007;44(11-12): 3875-3886.